\documentclass[11pt]{article}
\usepackage{epsfig}
\usepackage{amsfonts}
\usepackage{bbm}
\newcommand{\la}[1]{\label{#1}}
\newcommand{\be}{\begin{equation}}
\newcommand{\ee}{\end{equation}}
\newcommand{\ba}{\begin{eqnarray}}
\newcommand{\ea}{\end{eqnarray}}
\newcommand{\rmi}[1]{{\mbox{\scriptsize #1}}}
\newcommand{\fig}{Fig.~}
\newcommand{\figs}{Figs.~}
\newcommand{\eq}{Eq.~}
\newcommand{\se}{Sec.~}
\newcommand{\eqs}{Eqs.~}
\newcommand{\nr}[1]{(\ref{#1})}
\newcommand{\tr}{{\rm Tr\,}}
\newcommand{\nn}{\nonumber \\}
\newcommand{\hb}[1]{\,\hat{\!{\bar #1}}}
\newcommand{\fr}[2]{{\frac{#1}{#2}\,}}

\renewcommand{\vec}[1]{{\bf #1}}

\newcommand{\Ms}{M_1} 

\newcommand{\Nc}{N_{\rm c}}

\def\lsi{\raise0.3ex\hbox{$<$\kern-0.75em\raise-1.1ex\hbox{$\sim$}}}
\def\gsi{\raise0.3ex\hbox{$>$\kern-0.75em\raise-1.1ex\hbox{$\sim$}}}
\newcommand{\lsim}{\mathop{\;\lsi\;}}
\newcommand{\gsim}{\mathop{\;\gsi\;}}

\newcommand{\fe}{\rmi{f}}

\newcommand{\nF}[1]{n_\rmi{F{#1}}}

\newcommand{\rmii}[1]{{\mbox{\tiny\rm{#1}}}}
\newcommand{\re}{\mathop{\mbox{Re}}}
\newcommand{\im}{\mathop{\mbox{Im}}}
\newcommand{\bsl}[1]{\,\slash\!\!\!\!{#1}\,}
\newcommand{\msl}[1]{\,\slash\!\!\!{#1}\,}
\newcommand{\Tint}[1]{{\hbox{$\sum$}\!\!\!\!\!\!\!\int}_{\!\!\!\!\raise-0.9ex\hbox{$\scriptstyle{#1}$}}}
\newcommand{\f}{f} 

\newcommand{\ZZ}{{\mathbb{Z}}}

\newcommand{\hide}[1]{ }

\makeatletter \@addtoreset{equation}{section} \makeatother
\renewcommand{\theequation}{\arabic{section}.\arabic{equation}}
\makeatletter
\renewcommand\section{\@startsection {section}{1}{\z@}%
                                   {-5.5ex \@plus -1ex \@minus -.2ex}
                                   {2.3ex \@plus.2ex}%
                                   {\normalfont\large\bfseries}}
\renewcommand\subsection{\@startsection{subsection}{2}{\z@}%
                                     {-3.25ex\@plus -1ex \@minus -.2ex}%
                                     {1.5ex \@plus .2ex}%
                                     {\normalfont\normalsize\bfseries}}
\renewcommand\thesection {\@arabic\c@section}
\renewcommand\thesubsection   {\thesection.\@arabic\c@subsection}
\renewcommand{\@seccntformat}[1]{%
\csname the#1\endcsname.\hspace{1.0em}}
\makeatother

\begin{document}

\begin{titlepage}
\begin{flushright}
BI-TP 2008/02\\
NSF-KITP-08-39\\
\end{flushright}
\begin{centering}
\vfill

\noindent
{\Large{\bf Sterile neutrino dark matter as a consequence \\[1mm] 
of $\nu$MSM-induced lepton asymmetry}}

\vspace{0.8cm}

Mikko~Laine$^\rmi{a,b}$ and
Mikhail~Shaposhnikov$^\rmi{c}$

\vspace{0.8cm}

$^\rmi{a}${\em
Faculty of Physics, University of Bielefeld, 
D-33501 Bielefeld, Germany\\}

\vspace{0.3cm}

$^\rmi{b}${\em
Department of Physics, University of Oulu, 
FI-90014 Oulu, Finland\\}

\vspace{0.3cm}

$^\rmi{c}${\em
Institut de Th\'eorie des Ph\'enom\`enes Physiques, EPFL, 
CH-1015 Lausanne, Switzerland}

\vspace*{0.8cm}
 
\mbox{\bf Abstract}

\end{centering}

\vspace*{0.3cm}
 
\noindent 
It has been pointed out in ref.~\cite{asy} that in the  $\nu$MSM
(Standard Model extended by three right-handed neutrinos with masses
smaller than the electroweak scale), there is a corner in the
parameter space where CP-violating resonant oscillations among  the
two heaviest right-handed neutrinos continue to operate below the 
freeze-out temperature of sphaleron transitions, leading to a lepton 
asymmetry which is considerably larger than the baryon asymmetry. 
Consequently, the lightest right-handed (``sterile'')  neutrinos,
which may serve as dark matter, are generated through an  efficient
resonant mechanism proposed by Shi and Fuller~\cite{Shi:1998km}.  We
re-compute the dark matter relic density and non-equilibrium
momentum  distribution function in this situation with quantum field
theoretic methods  and, confronting the results with existing
astrophysical data, derive  bounds on the properties of the lightest
right-handed neutrinos.  Our spectra can be used as an input for
structure formation simulations  in warm dark matter cosmologies, for
a Lyman-$\alpha$ analysis  of the dark matter distribution on small
scales, and for studying  the properties of haloes of dwarf
spheroidal galaxies.
\vfill
\noindent
 

\vspace*{1cm}
 
\noindent
June 2008

\vfill

\end{titlepage}

%
\section{Introduction}

Ever since the experimental discovery of neutrino mass differences,
there  has been a compelling case for the existence of right-handed
neutrinos  in nature. It turns out, however, to be difficult to
determine  the parameters associated with them with any
precision.  Indeed, given that right-handed neutrinos are gauge
singlets, their Lagrangian contains explicit (Majorana) mass terms in
addition to the  usual Yukawa interactions. The known mass
differences only constrain certain combinations of the Yukawa
couplings and Majorana masses, so that the absolute scale of the
Majorana masses cannot be fixed from the existing data. 

Recently, it has been pointed out~\cite{Asaka:2005an,Asaka:2005pn}
that if the Majorana masses  are chosen to be significantly smaller
than has been the common  choice (this corner of the parameter space
was named $\nu$MSM,  for ``neutrino Minimal Standard Model''),  then
it appears possible
to find an amazingly complete description of the  main cosmological
mysteries that  cannot be explained within the Standard Model. 
Suppose that there are three generations of the right-handed
neutrinos,  like there are of all other fermions in the Standard
Model. Then  the lightest right-handed, or ``sterile'' neutrinos, 
with masses in the keV range,  might serve as (warm) 
dark matter~\cite{Dodelson:1993je,Shi:1998km}, 
\cite{Dolgov:2000ew}--\cite{Kishimoto:2008ic}\footnote{%
 Various cosmological and
 astrophysical phenomena related to the dark matter sterile neutrinos 
 have been discussed in refs.~\cite{astro}.
 }; 
the two  heavier right-handed neutrinos, with masses in the GeV range
and almost degenerate with each other, could account  simultaneously
for baryogenesis and the observed active neutrino mass 
matrix~\cite{Akhmedov:1998qx,Asaka:2005pn};  while a non-minimal
coupling of the Higgs  field in this theory to the Ricci scalar might
explain  inflation~\cite{Bezrukov:2007ep}. In fact it can be  argued
that the $\nu$MSM could be a good effective field theory  all the way
up to the Planck scale~\cite{Shaposhnikov:2007nj} (for a similar
argument in a related theory,  see ref.~\cite{Meissner:2006zh}).

On the quantitative level, though, it is non-trivial to realize all
of these possibilities within the $\nu$MSM. Consider the explanation of
dark matter by the lightest sterile neutrinos, for instance. There
are strong  experimental constraints from two sides: from the
non-observation  of an X-ray signal generated by the decay of the
dark  matter neutrinos on one  hand
(\cite{Boyarsky:2005us}--\cite{Boyarsky:2007ge} and references
therein),  and from structure formation simulations on the 
other~\cite{Hansen:2001zv}--\cite{Viel:2007mv}.  Combining these
experimental constraints with the results of theoretical 
computations of thermal dark matter production due 
to active-sterile mixing (the so-called 
Dodelson-Widrow mechanism)~\cite{Dodelson:1993je,Dolgov:2000ew, 
Abazajian:2001nj,Abazajian:2002yz,Abazajian:2005gj,als2}  
appears in fact to all but exclude  the warm dark matter
scenario~\cite{Seljak:2006qw}--\cite{Viel:2007mv},
\cite{Boyarsky:2007ay}\footnote{%
 Extending the $\nu$MSM by an extra scalar field allows
 for an additional mechanism for dark matter sterile neutrino 
 creation which evades these 
 limits~\cite{Shaposhnikov:2006xi,Kusenko:2006rh,Petraki:2007gq};
 another relaxation follows 
 if the reheating temperature after inflation is low 
 (in the MeV range)~\cite{Gelmini:2004ah,Yaguna:2007wi,Khalil:2008kp}. 
}. 

Such a negative conclusion is premature, however. First of all, the
structure formation simulations of 
refs.~\cite{Seljak:2006qw}--\cite{Viel:2007mv} assumed the spectrum
of the dark matter sterile neutrinos to be  {\em thermal}, with
possibly a modest shift of the average  momentum towards the
infrared, while in reality the deviations from the  Fermi-Dirac
distribution are substantial~\cite{Abazajian:2005gj,als2}. Second,
and even more importantly, it has been pointed out in 
ref.~\cite{asy} that in the framework of the $\nu$MSM it is  possible
to generate a large leptonic chemical potential  surviving down to
temperatures of a few hundred MeV. In this situation the  results of
the theoretical computation change 
dramatically~\cite{Shi:1998km,Abazajian:2001nj}, 
and it becomes easier to explain dark matter with sterile
neutrinos. 

The purpose of the present paper is to elaborate on the latter 
possibility. More precisely, we re-compute the dark matter relic 
density in this situation with the quantum field theoretic methods
introduced in refs.~\cite{als,als2}; analyse uncertainties related 
to unknown parameters and poorly known QCD phenomena; and compare 
with previous computations in the literature. Our main finding  is
that if lepton asymmetries in the range $n_{\nu_e}/s \gsim  
0.8\times 10^{-5}$ exist,  where $n_{\nu_e}$  refers to the asymmetry
in   electron-like neutrinos and $s$ is  the total entropy density,
then the $\nu$MSM  can indeed account for the observed dark matter
abundance. This bound can be consolidated
once structure formation simulations have been repeated
with the  non-equilibrium momentum distributions functions 
(``spectra'') of the sterile neutrinos that we derive.
In any case, asymmetries in the range 
$n_{\nu_e}/s \gsim  0.8\times 10^{-5}$ may be 
reachable in the so-called  {\bf Scenario IIa} of parameter values 
of ref.~\cite{asy}. Note that the 
largest possible asymmetry leading to successful 
Big Bang Nucleosynthesis corresponds to
a chemical potential $|\mu_L/T|\lsim 0.07$ at 
$T\sim 1$~MeV (95\% CL)~\cite{Kohri:1996ke,Dolgov:2002ab}, meaning
$n_{\nu_e}/s \lsim 2.5 \times 10^{-3}$ in our units (cf.\ appendix~A).
The maximal asymmetry which can be produced within the $\nu$MSM is
somewhat smaller, $n_{\nu_e}/s \lsim 0.7 \times 10^{-3}$~\cite{asy}.

It is appropriate to stress that even though our considerations are
naturally viewed as a sequel to the lepton asymmetry generated
{\em \`a la} ref.~\cite{asy}, from a practical point of view the
origin of the lepton asymmetry plays no role in the present analysis,
nor do  the parameters related to the two heaviest right-handed
neutrinos.   Indeed the only ingredient entering our computation is
the absolute  value of the lepton asymmetry which, as mentioned, we
parametrize through the ratio $n_{\nu_e}/s$. Recalling that the
observed baryon asymmetry is 
$n_B/s \simeq (0.9 - 1.0) \times 10^{-10}$~\cite{pdg,Komatsu:2008hk},
we hence need to assume a boost of some five orders of magnitude in
the leptonic sector.  Besides $\nu$MSM, another possible origin for
such an asymmetry could be the Affleck-Dine
mechanism~\cite{Affleck:1984fy}, if it takes  place below the
electroweak  scale and is based on a condensate producing many more
leptons than quarks.

Our presentation is organized as follows.  In \se\ref{se:general} we
generalize the formalism  of refs.~\cite{als,als2} to the
charge-asymmetric situation. The resulting equation  for the sterile
neutrino abundance is integrated numerically in \se\ref{se:abundance}, 
and the equation for the sterile neutrino  spectrum 
in \se\ref{se:spectrum}. We discuss the astrophysical  
consequences of these results in \se\ref{se:astro}, and conclude  
in \se\ref{se:concl}. In appendix~A we recall the relations of our
characterization of the lepton asymmetry, through $n_{\nu_e}/s$, to 
a number of other conventions appearing in the literature.  A reader
only interested in the phenomenological consequences  of our analysis
could start directly from \se\ref{se:astro}.

%
\section{Basic formalism}
\la{se:general}

Our starting point is the Lagrangian
\be
 \mathcal{L} = 
 \fr12 \bar{\tilde N_1} i \msl{\partial} \tilde N_1 
 - \fr12 M_1 \bar{\tilde N_1} \tilde N_1
 - F_{\alpha 1} \bar L_\alpha \tilde \phi\, a_R \tilde N_1
 - F_{\alpha 1}^* \bar{\tilde N_1} \tilde \phi^\dagger a_L L_\alpha
 + \mathcal{L}_\rmi{MSM}
 \;,  \la{LM}  
\ee
where $\tilde N_1$ are Majorana spinors, 
and the subscript ``1'' refers to the lightest right-handed neutrino; 
repeated indices are summed over; 
$M_1$ is the Majorana mass that we have chosen to be real in 
this basis; $L_\alpha$ are the
active lepton doublets; $F_{\alpha 1}$ are elements of a 
complex Yukawa matrix;  $\tilde\phi = i \tau_2 \phi^*$ is the
conjugate Higgs doublet; and $a_L \equiv (1-\gamma_5)/2$,  $a_R
\equiv (1+\gamma_5)/2$ are chiral projectors.

To compute the abundance of $N_1$ from first principles, we make a
number of basic assumptions, following refs.~\cite{als,als2}. First
of all, we restrict to temperatures below a few GeV,  implying that
the electroweak symmetry is broken:  $\langle \tilde \phi \rangle
\simeq (v/\sqrt{2},0)$,  where  $v\simeq 246$~GeV is the Higgs field
vacuum expectation value.  Second, we assume that  the mixing angles 
$\theta_{\alpha 1}^2 \equiv |M_D|_{\alpha 1}^2/M_1^2$,  where
$|M_D|_{\alpha 1} \equiv |v F_{\alpha 1}| /\sqrt{2}$,   are very
small, $\theta_{\alpha 1} \lsim 10^{-3}$.  Then it is sufficient to
restrict to the leading non-trivial order in a Taylor series in
$\theta_{\alpha 1}^2$. 

The third assumption concerns the flavour structure  of the lepton
asymmetry.  We will only consider the case when the asymmetries in all
active species  ($\nu_e$, $e_L$, $e_R$, $\nu_\mu$, $\mu_L$, $\mu_R$, 
$\nu_\tau$, $\tau_L$, $\tau_R$) are equal. Strictly speaking, this is
not satisfied in the $\nu$MSM, since the generation of lepton
asymmetries takes place when the reactions changing neutrino flavours
freeze out~\cite{asy}.  We make this assumption in order to keep the
discussion as simple as possible, and also because it yields the
most conservative constraints, leading to the largest sterile neutrino
abundance and  consequently weakening the X-ray bounds. 
We discuss the general formalism applicable 
in this setting in \se\ref{ss:muL}.

At the same time, there is no reservoir replenishing the lepton
asymmetry if a part of it is converted to right-handed neutrinos: 
the CP-violating reactions generating the asymmetry cease to take place 
at temperatures above a few GeV~\cite{asy}. Together with the third 
assumption this completely fixes the time evolution of the lepton 
asymmetry; this will be demonstrated in \se\ref{ss:BR1}.

The fourth and final assumption asserts that the density of
the right-handed neutrinos produced is below its equilibrium
value.  This assumption is necessary for the validity of 
the quantum field theoretic formulation of refs.~\cite{als,als2}; 
on the other hand, it may be violated in certain parts of 
the parameter space. In \se\ref{ss:BR2} 
we outline a phenomenological way to correct for a possible violation.

%
\subsection{Results in terms of a generic lepton asymmetry}
\la{ss:muL}

Under the assumption of a chemically equilibrated 
lepton asymmetry among the active species, 
the formal determination of the sterile
neutrino production rate proceeds almost exactly as in
ref.~\cite{als}, with the only difference that 
the density matrix of the  Minimal Standard
Model (MSM) now takes the form\footnote{%
  In general, different
  chemical potentials have to be introduced for different 
  leptonic flavours.} 
\be
 \hat \rho_\rmi{MSM} 
 = Z^{-1}_\rmi{MSM} \exp[-\beta (\hat H_\rmi{MSM} - \mu_L \hat L_\rmi{MSM})]
 \;. 
\ee 
Here  $\beta \equiv 1/T $;  $\mu_L\neq 0$ unlike in
ref.~\cite{als};  and $\hat L_\rmi{MSM}$ is the total lepton number
operator within the MSM, 
\be
 \hat L_\rmi{MSM} \equiv
 \int \!{\rm d}^3 \vec{x} \, 
 \sum_{\alpha = e,\mu,\tau}
 \Bigl[
    \hb{l}_{\alpha L} \,\gamma_0\, \hat{l}_{\alpha L}  
   + \hb{l}_{\alpha R} \,\gamma_0\, \hat{l}_{\alpha R}  
   + \hb{\nu}_{\alpha L} \,\gamma_0\, \hat{\nu}_{\alpha L}   
 \Bigr]
 \;,
\ee  
with 
$ 
 l_e \equiv e, l_\mu \equiv \mu, l_\tau \equiv \tau
$; and 
$\hat \psi_L \equiv a_L\hat \psi$, 
$\hat \psi_R \equiv a_R\hat \psi$.

According to ref.~\cite{als},
the phase space density $\f_1(t,\vec{q})$ of
right-handed neutrinos in either helicity state $s$, 
\be
 \f_1(t,\vec{q}) \equiv \sum_{s=1,2}
 \frac{{\rm d}
 N_1^{(s)}(t,\vec{x},\vec{q})}{{\rm d}^3 \vec{x}\,{\rm d}^3 \vec{q}} 
 \;, \la{n1_norm}
\ee
obeys the equation 
\be
  \biggl( \frac{\partial}{\partial t} - 
  H q_i \frac{\partial}{\partial q_i}\biggr) 
  \f_1(t,\vec{q}) 
  = R(T,\vec{q})
 \;. 
 \la{expansion}
 \la{kinetic}
\ee
Here $H$ is the Hubble parameter,  $H={\rm d}\ln a(t)/ {\rm d}t$, 
and $q_i$ are the spatial components of the physical momentum 
$\vec{q}$, defined in a local Minkowskian frame.  Repeating
the analysis of ref.~\cite{als} with $\mu_L \neq 0$, 
the source term reads
\be
 R(T,\vec{q})
 = \frac{1}{(2\pi)^3 q^0}
   \sum_{\alpha = 1}^{3}  
   |M_D|^2_{\alpha 1} 
  \tr\Bigl\{ 
  \bsl{Q}
  a_L 
   \Bigl[
    \nF{}(q^0 - \mu_L) \rho_{\alpha\alpha}(Q)  +
    \nF{}(q^0 + \mu_L) \rho_{\alpha\alpha}(-Q)  
   \Bigr]
  a_R
 \Bigr\} 
 \;, \la{master}
\ee
where 
$\nF{}(q) \equiv 1/[\exp(q/T)+ 1]$ is the Fermi distribution function; 
$\rho_{\alpha\alpha}$ is the spectral function related to  the
propagator of the active neutrino of generation $\alpha$;  and $Q$ is
the on-shell four-momentum of the right-handed neutrino,  i.e.\ $Q^2
= M_1^2$. Noting that the solution of  \eq\nr{kinetic} only depends
on $q\equiv |\vec{q}|$,  it can be written as~\cite{als2}
\be
 \f_1(t_0,q)= \int_{T_0}^\infty \! \frac{{\rm d}T}{T^3} \, 
 \frac{M_0(T)}{3 c_s^2(T)}
 R\biggl( {T},
 q \frac{T}{T_0}
 \left[\frac{h_\rmi{eff}(T)}{h_\rmi{eff}(T_0)}\right]^{\frac{1}{3}}
 \biggr)
 \;,
 \la{distribution}
\ee
where  
$
 M_0(T) \equiv M_\rmi{Pl} 
 \sqrt{
 {45} / {4\pi^3
 g_\rmi{eff}(T)} 
 }
$; 
$g_\rmi{eff}(T)$ 
parametrizes the energy density $e$ as 
$
 e \equiv 
 {\pi^2 T^4 g_\rmi{eff}(T)}/{30} 
$; 
and
$h_\rmi{eff}(T)$ 
parametrizes the entropy density $s$ as 
$
 s \equiv
 {2 \pi^2 T^3} h_\rmi{eff}(T) / {45}  
$.
Moreover, $c_s^2$ is the sound speed squared, given by
$
 1/{c_s^{2}(T)} = 
 3 + {T h'_\rmi{eff}(T)} / {h_\rmi{eff}(T)} 
$.
Note that in \eq\nr{distribution}, the chemical potential $\mu_L$ 
may be taken to depend on $T$ in any way, to be specified 
later on from physical considerations. 

To derive an expression for $\rho_{\alpha\alpha}$,  we follow the
steps in Sec.~3.2 of ref.~\cite{als}, but without assuming anything
about the symmetry properties of the active neutrino self-energy for
the moment.  The Euclidean propagator (cf.\ \eq(3.10) of
ref.~\cite{als}) then becomes
\be
 \Pi^E_{\alpha\alpha}(\tilde Q) = a_L 
 \frac{1}{- i \bsl{\tilde Q} + i \bsl{\tilde\Sigma}
 ( -\tilde Q)} a_R 
 = a_L\, \frac{i \bsl{\tilde Q} - i \bsl{\tilde\Sigma}(-\tilde Q)}
   {[\tilde Q - \tilde \Sigma(-\tilde Q)]^2}\, a_R
 \;. \la{prop}
\ee
We have left out the flavour indices from the active neutrino 
self-energy $\tilde\Sigma$ to compactify the notation somewhat,  and
the tildes are a reminder of Euclidean conventions. 

Carrying out the Wick rotation, we can transform \eq\nr{prop} into a
retarded Minkowskian propagator  (cf.\ \eq(3.11) of ref.~\cite{als}):
\be
 \Pi^R_{\alpha\alpha}(q^0,\vec{q}) = 
 \Pi^E_{\alpha\alpha}(-i q^0,\vec{q}) = 
  a_L \frac{-\bsl{Q} + \bsl{\Sigma}(-Q)}
 {Q^2 - 2 Q\cdot \Sigma(-Q) + \Sigma^2(-Q)} a_R
 \;. \la{retd}
\ee
Writing finally
$\Sigma(q^0\pm i 0^+,\vec{q}) \equiv 
\re \Sigma(Q) \pm i \im \Sigma(Q)$ 
and correspondingly
$\Sigma(-[q^0\pm i 0^+],-\vec{q}) = 
\re \Sigma(-Q) \mp i \im \Sigma(-Q)$, 
allows to obtain (cf.\ \eqs(3.1), (3.4) of ref.~\cite{als}) 
\ba
 \rho_{\alpha\alpha}(Q) & = & \frac{1}{2i} 
 \Bigl[ 
   \Pi^R_{\alpha\alpha}(q^0 + i 0^+,\vec{q}) - 
   \Pi^R_{\alpha\alpha}(q^0 - i 0^+,\vec{q})
 \Bigr]
 \\
 & = & a_L 
 \frac{
   - S_I(-Q)
      [ \bsl{Q} - \re \bsl{\Sigma}(-Q)]
   - S_R(-Q)
      \im\bsl{\Sigma}(-Q) 
  }
  {
   S_R^2(-Q) + S_I^2(-Q) 
  }
 a_R 
 \;,  \la{rhoQ}
\ea
where
\ba
 S_R(-Q) & \equiv & 
 [Q - \re \Sigma(-Q)]^2 - [\im\Sigma(-Q)]^2 
 \;, \la{SR} \\ 
 S_I(-Q) & \equiv & 
 -2 [Q - \re\Sigma(-Q) ] \cdot \im\Sigma(-Q) 
 \;. \la{SI} 
\ea
These expressions can trivially be written also for
$\rho_{\alpha\alpha}(-Q)$, and the results  can then be inserted into
\eq\nr{master}. The outcome constitutes a generalization of \eq(3.12)
of ref.~\cite{als}.

To compactify the resulting equations somewhat, we make the following
simplifications. We note, first of all, that  in the imaginary part
$\im\Sigma$, the chemical potential changes the thermal distribution
functions of the on-shell leptons that appear  in the intermediate
states. Given that we are interested in the case $\mu_L/T\ll 1.0$,
however, these changes are not important, and will  be ignored in the
following. Then we can assume that $\im\Sigma$ does  not get modified
by $\mu_L$ and, in particular, that $\im\Sigma(-Q) = \im\Sigma(Q)$, 
as is the case for $\mu_L = 0$. 

As far as $\re\Sigma$ is concerned, we note that its general 
structure can be written as 
\be
 \re \bsl{\Sigma}_{\alpha\alpha}(Q) = 
 \bsl{Q} a_{\alpha\alpha}(Q) + \msl{u} b_{\alpha\alpha}(Q)
 + \msl{u} c_{\alpha\alpha}(Q)
 \;,  \la{Restruct} 
\ee
where $u=(1,\vec{0})$.  We note that the function
$a_{\alpha\alpha}(Q)$ can be ignored, since  it is small compared
with the tree-level term $\bsl{Q}$.  On the other hand the latter
structures in \eq\nr{Restruct} do not appear at tree-level, and need
to be kept. We have separated two terms:  a function
$b_{\alpha\alpha}(Q)$ odd in $Q$, appearing already in the
charge symmetric situation, as well as a function  $c_{\alpha\alpha}(Q)$,
defined to be even in $Q$. The function $c_{\alpha\alpha}(Q)$ must be
proportional to the leptonic chemical potential (or leptonic net
number densities),  and it is this function which plays an essential
role in  the following. The explicit expression for
$c_{\alpha\alpha}(Q)$ can be extracted from ref.~\cite{ReSigmaold}; 
for $q^0\ll m_W$, we can work to first order in an expansion in
$1/m_W^2$, and then the result reads
\be
  c_{\alpha\alpha} = 3 \sqrt{2} G_F 
  (1 + 4 \sin^2\! \theta_\rmii{W}\,) n_{\nu_e}
  \;,  \la{c_simple}
\ee
where $G_F = g_w^2/4\sqrt{2} m_W^2$ is the Fermi constant and, 
as mentioned, we
assumed that all active leptonic densities are equal:
$
 n_{\nu_e} =
 n_{e_L} = n_{e_R} = n_{\mu_L} = ...~
$.\footnote{
 In general, 
 $
 c_{\alpha\alpha} =  
 \sqrt{2} G_F
 [ 
   (1 + 2 x_\rmii{W}\,) n_{l_{\alpha L}}
   -(1 - 2 x_\rmii{W}\,) 
   \sum_{\beta\neq\alpha }n_{l_{\beta L}}
   + 
   2 x_\rmii{W}\, \sum_{\beta} n_{l_{\beta R}}
   + 2 n_{\nu_\alpha}  
   + \sum_{\beta\neq \alpha} n_{\nu_\beta}
 ] 
 $,
 where $x_\rmii{W} \equiv \sin^2\! \theta_\rmii{W}$.  
}

With these simplifications, we can write 
\ba
 \bsl{Q} + \re \bsl{\Sigma} (Q) & \approx & 
 \bsl{Q} + \msl{u} ( b + c ) 
 \;, \\[1mm]
 \Bigl[ Q + \re\Sigma(Q) \Bigr]^2 & \approx & 
 M_1^2 + 2 q^0 (b + c) + (b + c)^2 
 \;, \\[2mm]
 \bsl{Q} - \re \bsl{\Sigma} (-Q) & \approx & 
 \bsl{Q} + \msl{u} ( b - c ) 
 \;, \\[1mm]
 \Bigl[ Q - \re\Sigma(-Q) \Bigr]^2 & \approx & 
 M_1^2 + 2 q^0 (b - c) + (b - c)^2 
 \;, 
\ea
where $b\equiv b_{\alpha\alpha}(Q)$, $c\equiv c_{\alpha\alpha}(Q)$.
Furthermore, all appearances of $\im\Sigma$ can be written  in terms
of the objects
\ba
 I_Q & \equiv & 
 \tr \Bigl[  \bsl{Q} a_L  \im \bsl{\Sigma}(Q) a_R \Bigr] 
 = 2\; Q \cdot \im\Sigma(Q)
 \;,  \la{IQ} \\
 I_u & \equiv & 
 \tr \Bigl[  \msl{u} a_L  \im \bsl{\Sigma}(Q) a_R \Bigr] 
 = 2\; u \cdot \im\Sigma(Q)
 \;. \la{Iu}
\ea
Note, in particular, that $\im \bsl{\Sigma}$ has a structure
analogous to \eq\nr{Restruct}, with one term proportional to 
$\bsl{Q}$ and another to $\msl{u}$, and that consequently even
$[\im\Sigma]^2$ can be written in terms of the structures in
\eqs\nr{IQ}, \nr{Iu}, as
\be
 \Bigl[\im\Sigma (Q) \Bigr]^2 
 = \frac{
   -I_Q^2 + 2 q^0 I_Q I_u - M_1^2 I_u^2 
   }
   {4 \vec{q}^2}
 \;.
\ee
Inserting these simplifications into \eqs\nr{master}, \nr{rhoQ}--\nr{SI}, 
we finally obtain
\ba
 && \hspace*{-1cm}
 R(T,{q}) 
  \approx  \frac{1}{(2\pi)^3 q^0}
   \sum_{\alpha = e,\mu,\tau}  
   {|M_D|^2_{\alpha 1}}
   \times
   \nn  
 &&  \hspace*{-5mm} \times   
 \biggl\{ 
 \nF{}(q^0 + \mu_L)
 \frac{
   2 S_I(Q) [M_1^2 + q^0 (b+c)]
   - S_R(Q) I_Q
  }
  {
  S_R^2(Q) + S_I^2(Q)  
  }
 + (c\to -c, \mu_L\to - \mu_L)
 \biggr\} 
 \;,  \la{master2}
\ea
where
\ba
 S_R(Q)  & = & 
 M_1^2 + 2 q^0 (b + c) + (b + c)^2 
 +\frac{
   I_Q^2 - 2 q^0 I_Q I_u + M_1^2 I_u^2 
   }
   {4 \vec{q}^2}
 \;, \la{SRQ} \\
 S_I(Q)  & = & 
 I_Q + (b + c) I_u
 \;.  \la{SIQ}
\ea

We remark that the production of the dark matter sterile neutrinos,
with masses in the keV range, takes place at temperatures  below a
few GeV (cf.\ \fig\ref{fig:Ts} below).  In this case, like already at
$\mu_L = 0$,   it is numerically a very good approximation to set the
term $I_u$ to zero, whereby  \eqs\nr{SRQ}, \nr{SIQ} simplify further.

Although the formulae given are valid beyond perturbation theory, 
a practical  application does make use of approximate
perturbative expressions for the functions $b, c$ and $I_Q$. It is
important to realise that  at the point of a resonance, where some of
the ``large'' terms ($M_1^2 + 2 q^0 b$ and $\pm 2 q^0 c$) cancel
against each other, the magnitude of the remainder is determined  by
higher order terms ($(b\pm c)^2$ and  $I_Q$).  A consistent treatment
to a certain order in perturbation theory would  hence require a
correspondingly precise (2-loop) determination of  the large terms $2
q^0 b, 2 q^0 c$. At the same time,  in a practical application we are
not sitting precisely at the point of a resonance, but  integrate
over its contribution, so that for instance a slight  misplacement of
the precise temperature at which the resonance  takes place plays
little role. Consequently, we continue to  use 1-loop expressions for
the functions $b$ and $c$ throughout\footnote{%
 For all quantities not specified explicitly in this
 section, we use the expressions given in ref.~\cite{als}.  
 }.  
Nevertheless, as we will see,
the fact that at the point of the resonance, $\im\Sigma$ plays a role
also in the denominator,  will imply that a large $\im\Sigma$ can
also lead to a decreased abundance, in contrast to the case of
non-resonant production,  where $\im\Sigma$ essentially only plays a
role in the numerator.  For a recent discussion of various
resonance-related phenomena, see ref.~\cite{db}.

To be more quantitative about the role of the resonance,  we can work
out its contribution to the  production rate semi-analytically. To a
good approximation,  the resonance is at the point where the function
\be
  \mathcal{F}(T) \equiv M_1^2 + 2 q^0 (b-c)
\ee
vanishes
(this comes from the latter term in \eq\nr{master2},
after the insertion of \eq\nr{SRQ}). 
Around this point, the production  rate in \eq\nr{master2} can be
approximated as
\ba
 R(T,{q}) 
  & \approx & \frac{\nF{}(q^0-\mu_L)}{(2\pi)^3 q^0}
   \sum_{\alpha = e,\mu,\tau}  
   {|M_D|^2_{\alpha 1}}
 [M_1^2 + q^0 (b-c)]  
 \frac{
   2 I_Q 
  }
  {
  \mathcal{F}^2(T) + I_Q^2  
  }
 \\ 
 & \approx &  
\frac{\nF{}(q^0-\mu_L)}{(2\pi)^2 2 q^0}
   \sum_{\alpha = e,\mu,\tau} 
   {|M_D|^2_{\alpha 1}}
 \, M_1^2 \,
 \delta(\mathcal{F}(T))
 \;,  \la{master3}
\ea
where we made use of the fact that $I_Q$ is very small,  in
order to identify a representation of the Dirac delta-function. 
Inspecting the expressions for $b$ and $c$, the function
$\mathcal{F}$ is positive at very low temperatures (because $M_1^2$
dominates)  and at very large temperatures (because $b$ dominates),
but for  sufficiently large $n_{\nu_e}/s$ and sufficiently small
$q/T$,  the term $c$ overtakes the others at intermediate
temperatures; there are then two zeros of $\mathcal{F}$, and it turns
out to  be the lower among these that gives the dominant
contribution.  We denote the corresponding temperature by $T_R$.
\eq\nr{distribution} can  now be integrated, to yield
\be
  \f_1(t_0,q)\approx 
   \sum_{\alpha = e,\mu,\tau} 
   {|M_D|^2_{\alpha 1}}
  \left.
  \frac{1}{T_R^3}
  \frac{M_0(T_R)}{3 c_s^2(T_R)}
 \frac{\nF{}(q^0-\mu_L)}{(2\pi)^2 2 q^0}
    \frac{M_1^2}
         {|\mathcal{F}'(T_R)|} 
	\right|_{\mathcal{F}(T_R) = 0}   
  \;. \la{resonance}
\ee 

%
\subsection{Time evolution of the lepton asymmetry}
\la{ss:BR1}

The main formula of the previous section, \eq\nr{master2}, 
depends on the leptonic chemical potential, $\mu_L$, and on 
the lepton asymmetry, $n_{\nu_e}$, the two of which are related 
through \eq\nr{n_vs_mu}. However, the dependence of these quantities on 
the time (or temperature) has been left open. We now need 
to insert further physical input in order to fix this dependence.  

It is important to realize, first of all, that no reservoir 
exists for the lepton asymmetry: as explained in ref.~\cite{asy}, 
the lepton asymmetry was generated by CP-violating processes active 
at temperatures around a few GeV, which subsequently ceased to operate. 

Second, the mass of the lightest right-handed neutrino, $M_1$, is
much below the temperature, $M_1 \ll T$. Therefore lepton asymmetry
violating processes, whose rate is proportional to $M_1^2$, can 
to a very good approximation be neglected. In other words, 
dark matter sterile neutrinos and the active leptons  
can be characterized by a conserved quantity, 
which we may call the total lepton number. 
In fact, this physics is effectively
already built in in \eq\nr{master}, which shows that the rate 
for generating any of the two sterile neutrino states is a sum of 
two terms, with opposite chemical potentials appearing in them, 
as is appropriate for ``particles'' and ``anti-particles''. 

As a result of these two facts, the resonant transitions from 
active to sterile neutrinos, or more precisely the C-odd part 
of \eq\nr{master},  
cause the original asymmetry to get 
{\em depleted}. If the resonance is very effective, the depletion 
is fast and thereby rapidly terminates the resonance phenomenon. 

To be more quantitative, we make the (optimistic) assumption 
that the flavour and chirality changing processes within 
the active generations
are fast enough to stay in thermal equilibrium. There is then a 
reservoir of nine spin-1/2 degrees of freedom (three generations, 
each with left-handed neutrinos and both-handed charged leptons)
converting to sterile neutrinos. Denoting the two terms in 
\eq\nr{master} by $R_-$ and $R_+$, respectively, 
\eq\nr{expansion} can then be split and subsequently
completed into a closed set of three equations 
(we also adopt an ansatz removing the terms proportional 
to the Hubble parameter from \eq\nr{kinetic}):
\ba
 \frac{{\rm d}}{{\rm d}t} \f_-\Bigl(t,q(t_0) \frac{a(t_0)}{a(t)}\Bigr)
 & = & R_-\Bigl(T,q(t_0) \frac{a(t_0)}{a(t)}\Bigr)
  \;, \la{expansion_2a} \\
 \frac{{\rm d}}{{\rm d}t} \f_+\Bigl(t,q(t_0) \frac{a(t_0)}{a(t)}\Bigr)
 & = & R_+\Bigl(T,q(t_0) \frac{a(t_0)}{a(t)}\Bigr)
  \;, \la{expansion_2b} \\
 \frac{{\rm d}}{{\rm d}t}
  \Bigl\{9\, a^3(t) n_{\nu_e}(t) \Bigr\} 
  &  = &  {a^3(t)} \int \! {\rm d}^3\vec{q} \, 
  \Bigl[ R_+(T,q) - R_-(T,q) \Bigr] 
  \;, \la{expansion_2c}
\ea
where the dark matter spectrum $\f_1$ is now 
represented by the sum $\f_1(T,q) = \f_{-}(T,q) + \f_{+}(T,q)$.
The structure of these equations is such that the total 
lepton charge in a comoving volume, 
\be
 {L}_\rmi{tot} \equiv a^3(t)
 \Bigl\{  9\, n_{\nu_e}(t) +
  \int \! {\rm d}^3\vec{q} \, 
  \Bigl[
  \f_-(t,q) - \f_+(t,q)
  \Bigr]
 \Bigr\}
 \;, 
\ee
indeed remains conserved, as must be the case for $M_1\to 0$. 

Note that within the approximation of \eq\nr{master3}, the term
$R_+$ could be omitted from \eqs\nr{expansion_2a}--\nr{expansion_2c}, 
which would simplify the system somewhat. Another practical simplification
is to solve the equations in terms of the temperature rather 
than the time, as we already did in \eq\nr{distribution}.

A rough estimate for when the depletion has a substantial 
impact can be obtained as follows. If {\em all} of the 
original lepton asymmetry converts to sterile neutrinos, 
then $n_{1}/s \ge 9 n_{\nu_e}/s$, where $s$ is the total 
entropy density,  and 
\be
 n_1(t_0) \equiv \int \! {\rm d}^3\vec{q} \, \f_1(t_0,\vec{q})
 \;. \la{na1} 
\ee 
Therefore the depletion is substantial if 
$n_{\nu_e}/s \lsim n_{1}/9 s$. Evaluating the right-hand 
side of this inequality for the case that sterile neutrinos account 
for all of dark matter, \eq\nr{Ya1_bound}, we get
$
 {n_{\nu_e}} / {s} \lsim 
 { 4.0 \times 10^{-4}} \times { \mbox{keV} } / { 9 M_1 }
$, i.e.
\be
 \frac{ M_1 }{ \mbox{keV} } \lsim 45 \, 
 \biggl( 10^6 \frac{n_{\nu_e}}{s} \biggr)^{-1}
 \;.
\ee
In other words, for a small initial asymmetry, the depletion has 
a significant impact at all the masses, while for a large initial 
asymmetry, the effect of the depletion is subdominant (because
there is more to deplete), unless the mass is small.

%
\subsection{Back reaction and equilibration}
\la{ss:BR2}

The derivation of the formulae that 
our work is based upon, \eqs\nr{expansion_2a}--\nr{expansion_2c}, 
contains the assumption that the particles produced do not 
thermalize, i.e.,\ that their density remains below the equilibrium 
value at all times. Let us investigate the validity of this assumption. 

It is relatively easy to establish that the {\em total} number density
of the sterile neutrinos produced 
does remain significantly below the 
equilibrium value. Indeed, the density of the sterile neutrinos is 
constrained from above by \eq\nr{Ya1_bound}, and consequently  
\ba
 \frac{n_{1}(T_0)}{n_\rmi{eq}(T_0)} & \lsim &
 4.0\times 10^{-4} \frac{s(T_0)}{n_\rmi{eq}(T_0)} 
 \frac{\mbox{keV}}{M_1} 
 \quad \approx \quad 
 9.64 \times 10^{-4} \, h_\rmi{eff}(T_0) \, \frac{\mbox{keV}}{M_1} 
 \nn & \lsim & 
 0.072 \, \frac{\mbox{keV}}{M_1} 
 \;, \la{na1_abs}
\ea
where we inserted  $n_\rmi{eq}(T_0) = 3
\zeta(3) T_0^3/2\pi^2$ (cf.\ \eq\nr{Delta_def})
as well as 
$h_\rmi{eff}(T_0) \lsim 75$ corresponding to $T_0
\lsim 1$~GeV~\cite{pheneos}. Thereby the lightest sterile neutrinos
appear indeed to be out of equilibrium in the whole range 
$M_1/\mbox{keV} \ge 0.1$ that we are interested in. 

It must be realized, however, that the inequality \nr{na1_abs}
is not sufficient to guarantee the absence of problems. Indeed, 
the spectrum of the sterile neutrinos produced is strongly tilted
towards the infrared~\cite{Shi:1998km,Abazajian:2001nj}. 
Given that we are considering fermions, it must then be checked 
that the Pauli exclusion principle is not violated at small momenta.  
Although an exact quantum field theoretic treatment would guarantee
this automatically, the assumption of a non-thermal result in the 
derivation of \eq\nr{master2} means that this consideration now
enters as an additional ingredient. We refer to the dynamics that 
prevents an excessive growth of the fermionic density at small 
momenta as ``back reaction''.

Motivated by Boltzmann equations, and recalling our normalization 
(cf.\ \eq\nr{n1_norm}), we expect that the way 
that back reaction works is to modify the source terms 
for the distribution functions $\f_-, \f_+$ 
(cf.\ \eqs\nr{master}, \nr{expansion_2a}, \nr{expansion_2b}) by replacing 
\be
 \frac{\nF{}(q^0 - \mu_L)}{(2\pi)^3} \to 
 \frac{\nF{}(q^0 - \mu_L)}{(2\pi)^3} - \f_-  
 \;, \quad 
 \frac{\nF{}(q^0 + \mu_L)}{(2\pi)^3} \to 
 \frac{\nF{}(q^0 + \mu_L)}{(2\pi)^3} - \f_+  
 \;.  \la{non-lin}
\ee
However, since this recipe would be purely phenomenological at this stage, 
and since the resulting equations are quite difficult to solve 
numerically\footnote{%
  If the spectra $\f_\mp$ do {\em not} appear in 
  $R_\mp$ on the right-hand side, 
  we can integrate over momenta in \eqs\nr{expansion_2a}, \nr{expansion_2b}, 
  to obtain a coupled set of ordinary differential equations for integrated
  densities; on the contrary, 
  if the spectra {\em do} appear
  in $R_\mp$ on the right-hand side, 
  modes with different momenta $q$ couple to each other 
  and need to be solved simultaneously, which 
  makes the problem significantly harder.
  }, 
we follow a simpler 
approach in the following. Indeed, we first solve 
\eqs\nr{expansion_2a}--\nr{expansion_2c} without back reaction, 
yielding the distribution functions which we denote by 
$
 \f_-^{(0)}
$, 
$
 \f_+^{(0)}
$. 
Subsequently, we construct the approximants
\be
 \f_\mp \simeq 
 \frac{\frac{\nF{}(q^0 \mp \mu_L)}{(2\pi)^3} \cdot \f_\mp^{(0)} }
 {\frac{\nF{}(q^0 \mp \mu_L)}{(2\pi)^3} + \f_\mp^{(0)} }
 \;.
 \la{iter}
\ee
This amounts to a rough iterative solution of the
structures suggested by \eq\nr{non-lin}; guarantees that 
$\f_\mp$ never exceed the equilibrium distributions 
$\nF{}(q^0\mp \mu_L)/(2\pi)^3$; and, for 
$\f_\mp \ll \nF{}(q^0\mp \mu_L)/(2\pi)^3$,  
yields the correct result $\f_\mp = \f_\mp^{(0)}$. 

In order to estimate the practical importance of 
the back reaction, we have determined our main 
observables (cf.\ Secs.~\ref{se:abundance}, \ref{se:spectrum}) 
for a number of parameter values both from 
$\f_\mp^{(0)}$ and $\f_\mp$. We return to the 
corresponding error estimates in connection with 
the numerical data. 

%
\section{Sterile neutrino abundance}
\la{se:abundance}

The task now is to evaluate the integral in \eq\nr{distribution}, 
with the lepton asymmetry evolved according to \eq\nr{expansion_2c}, 
producing the distribution function $\f_1(t_0,q)$, with $q\equiv |\vec{q}|$;
and then to integrate over $\vec{q}$, to get the total number density $n_1$
(the order of the integrations over $T$ and $q$ can 
of course be interchanged).  
We choose $t_0$ to be the time corresponding to $T_0 = 1$~MeV, 
below which active neutrinos start to decouple.
In order to present the numerical results, 
we start by  introducing some further notation.

First of all,  it is conventional to express the mixing angle as
$\sin^2 \! 2 \theta_{\alpha 1}$. For the  very small  Yukawa
couplings that we are interested in, it is an excellent approximation
to write 
\be
 \sin^2 \! 2 \theta_{\alpha 1} = 4 \theta^2_{\alpha 1} 
 = 4 \frac{|M_D|_{\alpha 1}^2}{M_1^2}
 \;. \la{thethe}
\ee
Second, we introduce a total mixing angle as 
\be
 \sin^2 \! 2 \theta \equiv \sum_{\alpha = e,\mu,\tau} 4 \theta_{\alpha 1}^2 
 \;, \la{sin_sum}
\ee 
which is the quantity appearing in the X-ray constraints
to be discussed below (cf.\ \fig\ref{fig:exclusion}). 

Now, we can write the total right-handed neutrino density as
$
 n_1(t_0) 
 \equiv \sum_{\alpha = e, \mu, \tau} n_{\alpha 1}(t_0)
$, 
where $n_{\alpha 1}$ is  the contribution from active flavour
$\alpha$ to the  dark matter abundance.  This contribution can
conveniently  be characterized through the yield parameter
$
 Y_{\alpha 1} \equiv {n_{\alpha 1}}/{s}
$.
The corresponding relative energy fraction is
$
 \Omega_{\alpha 1} \equiv {M_1 n_{\alpha 1}} / {\rho_\rmi{cr}} = 
          {M_1 Y_{\alpha 1}}/({\rho_\rmi{cr}/s})
$.
Inserting 
$
 \rho_\rmi{cr}/s \approx 3.65 \times 10^{-9} h^2 \, \mbox{GeV}
$ 
from Particle Data Group~\cite{pdg}, 
and noting that $\Omega_{\alpha 1} h^2$ can amount to 
at most the experimentally known dark-matter density,
$
  \Omega_{\rm dm} h^2 = 0.1143 \pm 0.0034 
$ (68\% CL)~\cite{Komatsu:2008hk}, 
we obtain an upper bound on $Y_{\alpha 1}$: 
\be
 Y_{\alpha 1} \lsim 4.0 \times 10^{-4}  \times \frac{ \mbox{keV} }{ M_1 }
 \;.
 \la{Ya1_bound}
\ee
Since $Y_{\alpha 1}$ depends monotonously (though non-linearly, 
because of the depletion discussed in \se\ref{ss:BR1})
on $\sin^2 \! 2 \theta_{\alpha 1}$, this equation yields
an upper bound on the mixing angle.

Following ref.~\cite{als2}, we concentrate particularly on two
flavour structures. In the non-resonant case,  for a fixed mixing 
angle, a hierarchy $Y_{e1} >
Y_{\mu 1} > Y_{\tau 1}$ can be observed~\cite{als2},  because heavier
scatterers suppress the production rate. The largest abundance, and
the most stringent upper  bound on $\sin^2\! 2 \theta$, is then
obtained with
\begin{eqnarray}
  | M_D |_{e 1} \neq 0\,,~~
  | M_D |_{\mu 1} = | M_D |_{\tau 1} = 0 \;
  \qquad\qquad\mbox{``case 1''}
  \;,
  \la{case1}
\end{eqnarray}
while the smallest relic abundance 
and the weakest upper bound on $\sin^2\! 2 \theta$
is obtained when
\begin{eqnarray}
  | M_D |_{\tau 1} \neq 0\,,~~
  | M_D |_{e 1} = | M_D |_{\mu 1} = 0 \;
  \qquad\qquad\mbox{``case 2''}
  \;.
  \la{case2}
\end{eqnarray}
As already mentioned, in the resonant case the roles of what leads to
the  strongest and weakest upper bound may get interchanged, because
the structure of \eq\nr{master2} is fairly complicated, but it
appears that these two cases should still capture the most extreme
possibilities.

To integrate \eqs\nr{distribution}, \nr{expansion_2c} 
in practice, we set  the upper
limit of the $T$-integration  to $T_\rmi{max} = 4$~GeV where $R(T,q)$
is vanishingly small,  and the lower limit to $T_0 = 1$~MeV. 
We first integrate over ${q}$, and then evolve $n_{\alpha 1}$, 
$Y_{\alpha 1}$ and $n_{\nu_e}/s$ through coupled ordinary 
differential equations in $T$. This is repeated with several 
$\sin^2 \! 2 \theta$ in order to find the value satisfying
the constraint in \eq\nr{Ya1_bound}.  
Our numerical
implementation follows that in ref.~\cite{als2}.  In particular, as
mentioned above, $\im\Sigma$ can be evaluated just as in the case
without a lepton asymmetry. At the same time, the existence of a
narrow resonance does make the integrations  over $q$ and $T$ much
more demanding than in the charge-symmetric case; 
most importantly, the resolution in the $q$-direction
needs to be significantly increased. 

We also remark that at small $M_1/$keV and large asymmetries, 
a direct numerical integration becomes increasingly
difficult, but at the same time \eq\nr{resonance} becomes 
more accurate. However, 
the asymmetry gets rapidly depleted in this regime, so that 
in fact  \eq\nr{resonance} is a good approximation only at 
the early stages of the resonance. We have found that a workable 
method is to approximate $R$ as a sum of a C-odd and C-even
part; the C-odd part is approximated by  \eq\nr{resonance}, while
the C-even part, which dominates for small asymmetries, is 
approximated by the full $R$ from ref.~\cite{als2}, 
with $\mu_L$ equal to zero. We have checked that the results
obtained this way extrapolate, within our resolution, 
to the ``exact'' results which
can be reliably determined at large masses, $M_1 \gsim 10$~keV.

\begin{figure}[t]


\centerline{%
\epsfysize=7.5cm\epsfbox{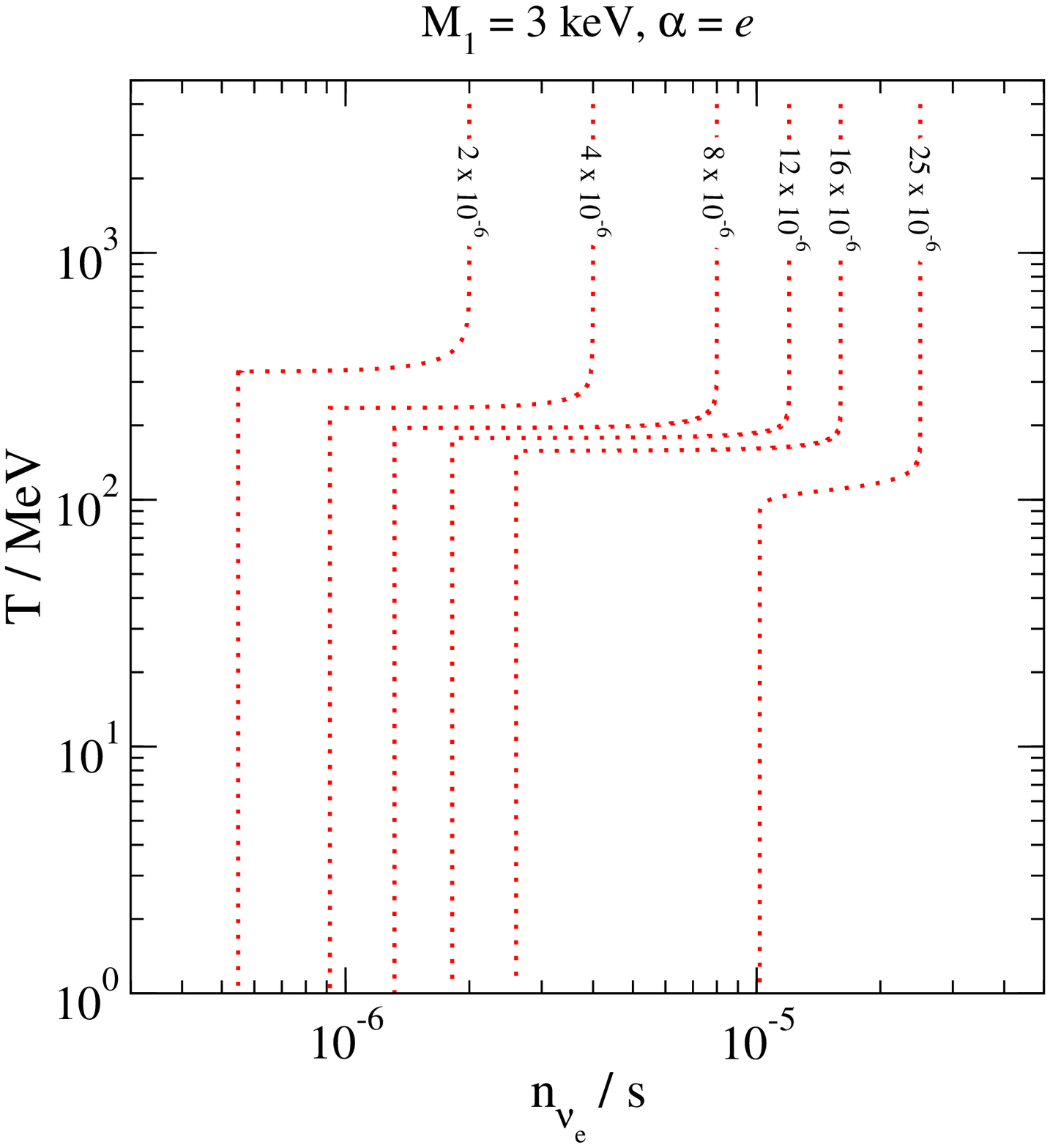}%
~~~\epsfysize=7.5cm\epsfbox{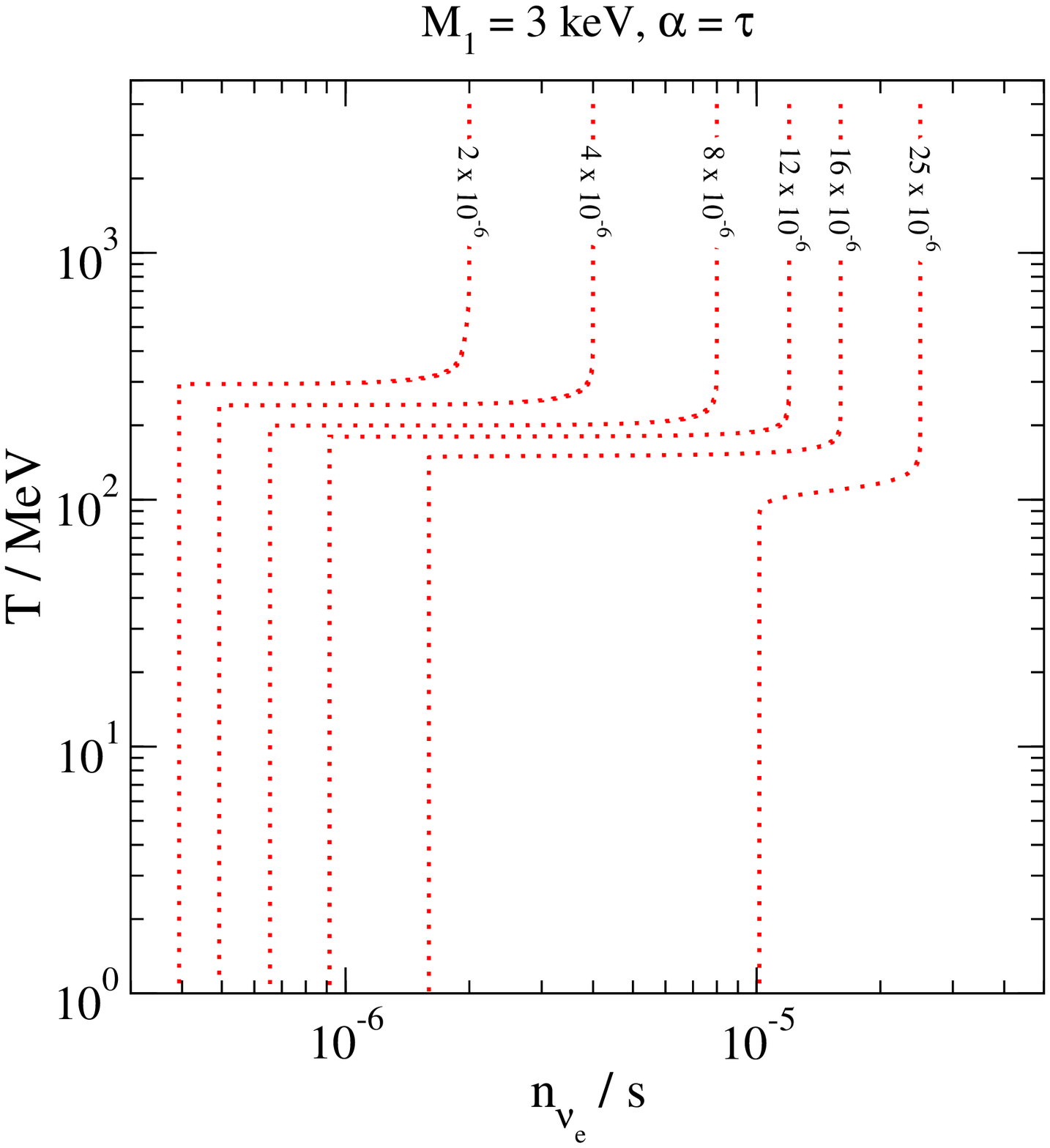}%
}

\caption[a]{\small Examples of the 
$T$-evolution of the lepton asymmetry $n_{\nu_e}/s$
(cf.\ \se\ref{ss:BR1}), for a fixed $M_1 = 3$~keV. 
Left: $\alpha = e$. Right: $\alpha = \tau$.
Note that our results differ even qualitatively from 
ref.~\cite{Kishimoto:2008ic} where the asymmetry
crosses zero at some temperature.
} 
\la{fig:Tevol}
\end{figure}
 
In \fig\ref{fig:Tevol} we show examples of the evolution of the
lepton asymmetry for various initial values.  It can be observed
that the resonance is quite narrow, and quite effective; in 
particular, for $n_{\nu_e}/s \lsim 10^{-6}$, most of the 
initial lepton asymmetry is rapidly converted to sterile neutrinos, 
so that the resonance becomes ineffective. Sterile neutrinos are
then dominantly produced thermally, like in ref.~\cite{als2}, 
and the mixing angle needs to be large. If the lepton asymmetry
reservoir were smaller, for instance with three components rather
than nine, then the depletion would be even more rapid. 

\begin{figure}[t]


\centerline{%
\epsfysize=7.5cm\epsfbox{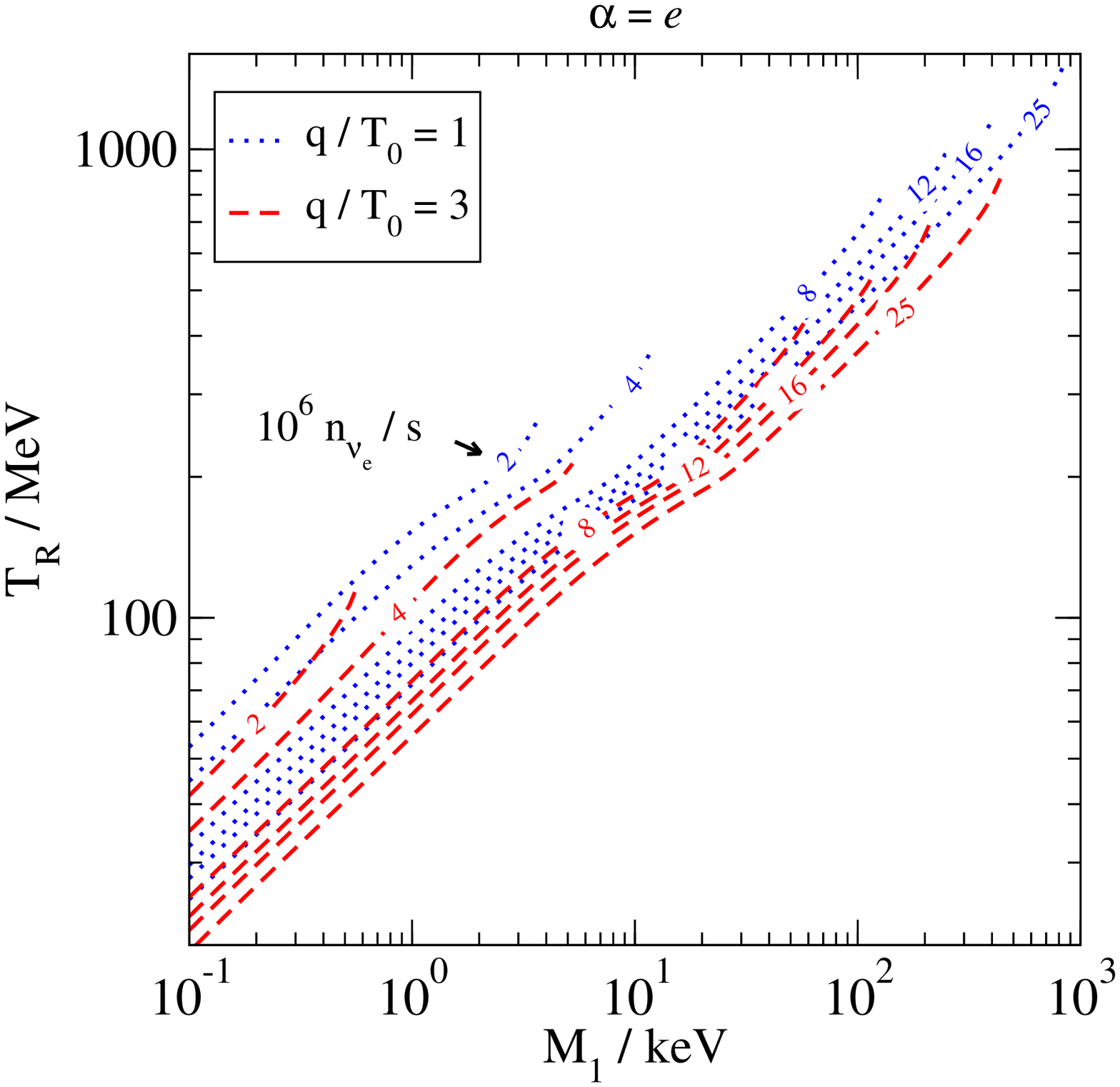}%
~~~\epsfysize=7.5cm\epsfbox{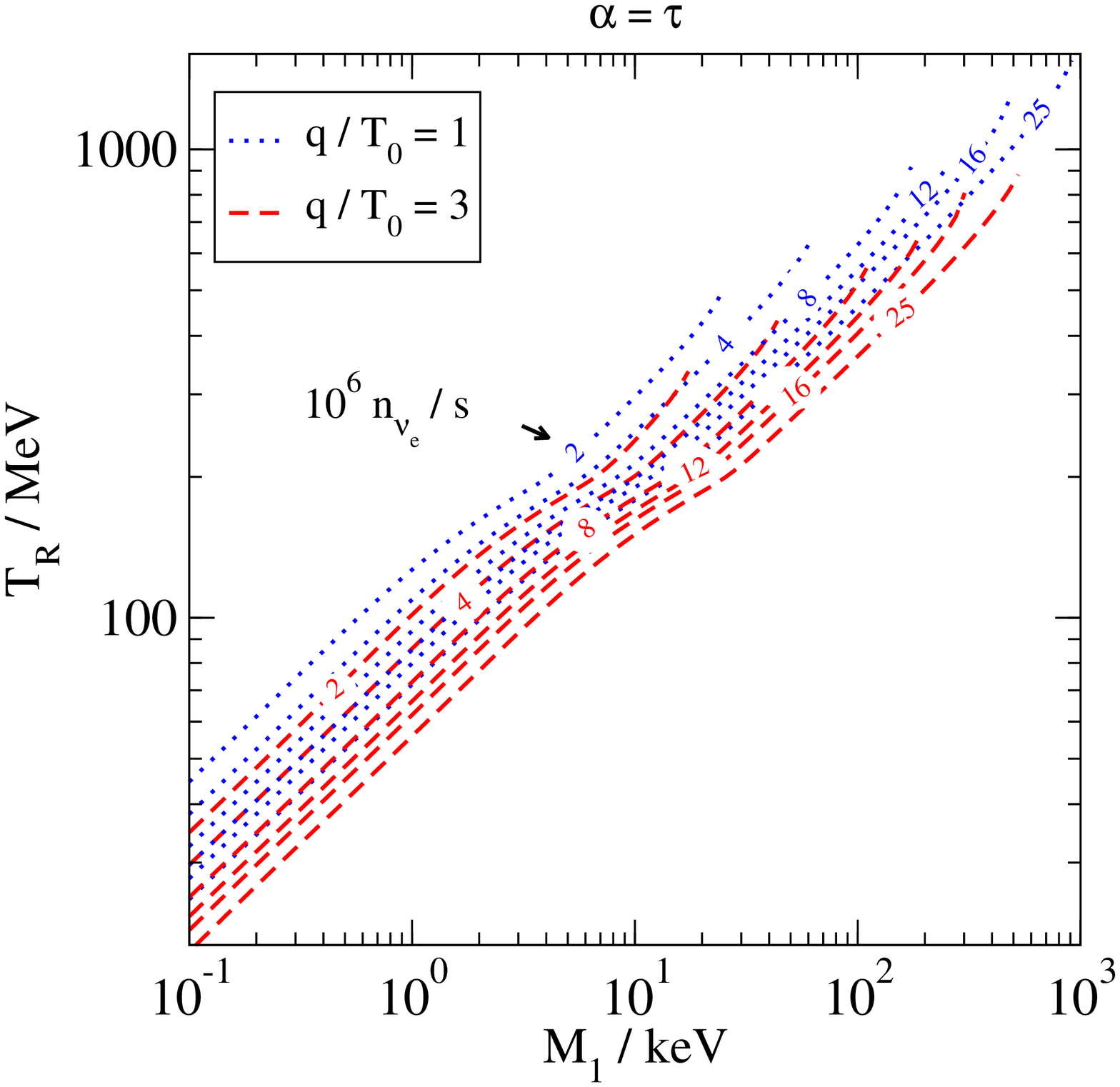}%
}

\caption[a]{\small The resonance temperature corresponding to
\eq\nr{resonance}, for the modes $q/T_0 = 1$ and $q/T_0=3$,  with
$T_0 = 1$~MeV. Left: $\alpha = e$. Right: $\alpha = \tau$. It is seen
that, for a given $M_1$,  the resonance first affects the smallest
values of $q/T_0$, and that the resonance extends to larger $M_1$
with increasing asymmetry (the asymmetry is indicated in units of
$10^6 n_{\nu_e}/s$ on top of the curves).} 
\la{fig:Ts}
\end{figure}

In \fig\ref{fig:Ts} we show the resonance temperatures (where
existent) for two momenta and various asymmetries, as a function
of the mass $M_1$. We note that for $M_1$ of a few keV, the 
production peaks at temperatures very close to the QCD crossover.
This introduces severe hadronic uncertainties to the results, 
as will be discussed below. 

\begin{figure}[t]


\centerline{%
\epsfysize=9.0cm\epsfbox{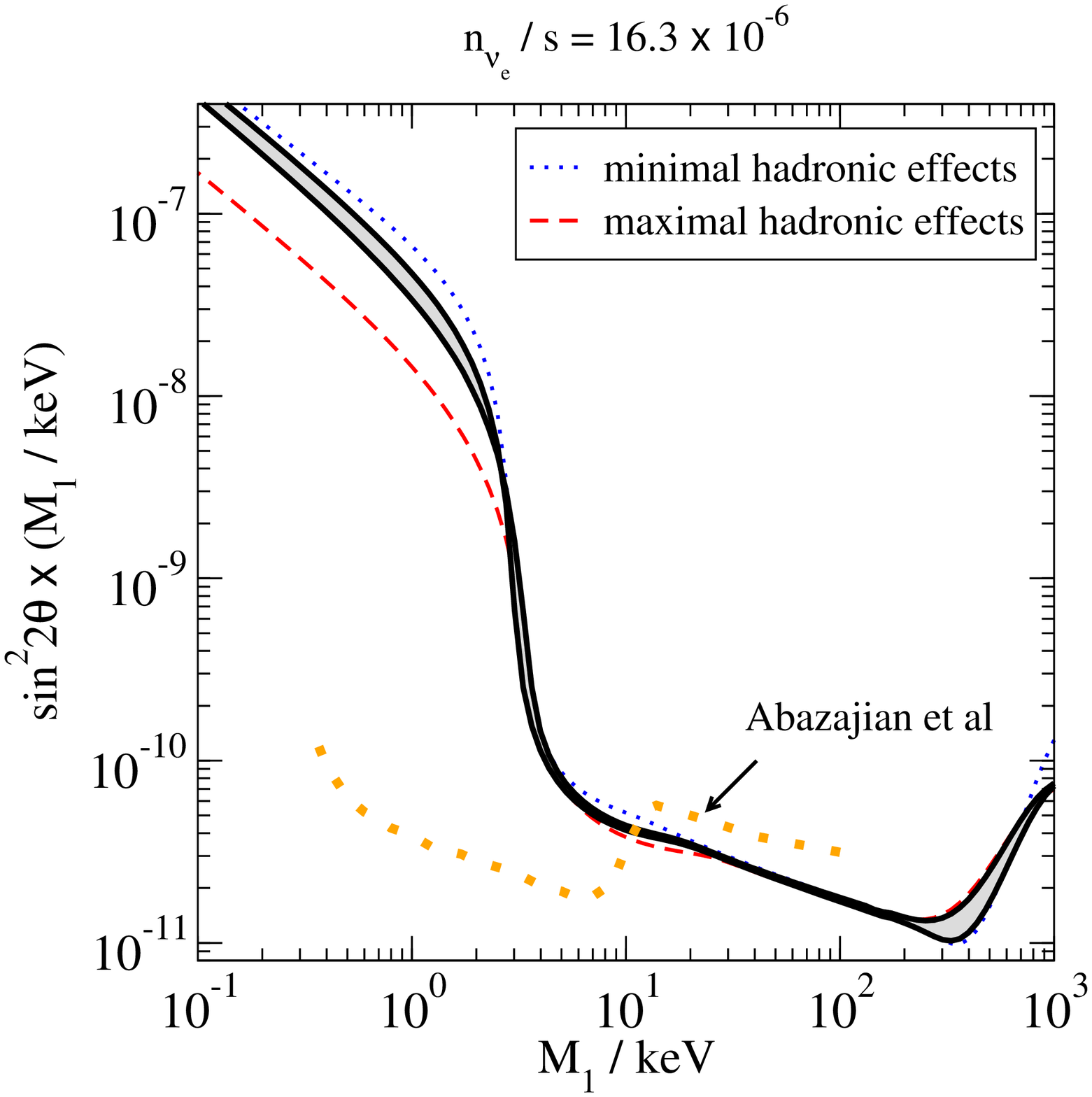}%
}

\caption[a]{\small 
The parameter values that, according to our
theoretical computation, 
lead to the correct dark matter abundance 
in the Shi-Fuller scenario~\cite{Shi:1998km};
if additional sources are present, $\sin^2\!2\theta$ must lie 
{\em below} the curves shown (cf.\ \eq\nr{Ya1_bound}).  For better visibility, 
the results have been multiplied by $M_1 / $keV.
The grey region between
case 1 (lower solid line on the left, upper solid line in the middle 
and on the right) 
and case 2 (other solid line) corresponds to different 
patterns of the active-sterile mixing angles,
cf.\ \eqs\nr{case1}, \nr{case2}.
The dotted and dashed lines  
correspond to one of these limiting
patterns with simultaneously 
the uncertainties from the equation-of-state and from hadronic scatterings 
set to their maximal values. The thick dotted line marked
with ``Abazajian et al'' shows
the result in \fig1 of ref.~\cite{Abazajian:2006yn}
(the case $L = 0.003$).
} 
\la{fig:exclusion_th}
\end{figure}

In \fig\ref{fig:exclusion_th} we show the upper bound 
on the mixing angle following from \eq\nr{Ya1_bound}, 
for $n_{\nu_e}/s = 16.3\times 10^{-6}$. This value has been  
chosen in order to allow for a comparison with fig.~1 of
ref.~\cite{Abazajian:2006yn}. It can be seen that at large
masses, $M_1 \gsim 3$~keV, the general order
of magnitude of our result is  remarkably close to the result of
ref.~\cite{Abazajian:2006yn}, despite the fact that this work was
using a simplified kinetic equation and  different approximations. 
On the other hand, at small masses, where the depletion discussed
in \se\ref{ss:BR1} is effective, the results are dramatically different. 

In \fig\ref{fig:exclusion_th} we have also considered the effects
of two sources of hadronic uncertainties: from
the  equation-of-state, which is defined to correspond  to a 20\%
rescaling of the pseudocritical temperature of the QCD crossover; 
and from hadronic scatterings, which is defined to 
correspond to evaluating the hadronic contributions to the vector and
axial current spectral functions with  {\em non-interacting} quarks,
and an effective number  $\Nc = 0$  or $\Nc = 3$ 
of colours. For justification and more details on this
phenomenological but  nevertheless conservative procedure, we refer
to \se5 of ref.~\cite{als2}. For plotting the dashed and dotted lines
in \fig\ref{fig:exclusion_th}, we have simultaneously set both
uncertainties to their maximal values. It is seen that the resulting
error depends strongly on parameters, but can be as large as 50\%.

\begin{figure}[t]


\centerline{%
\epsfysize=7.5cm\epsfbox{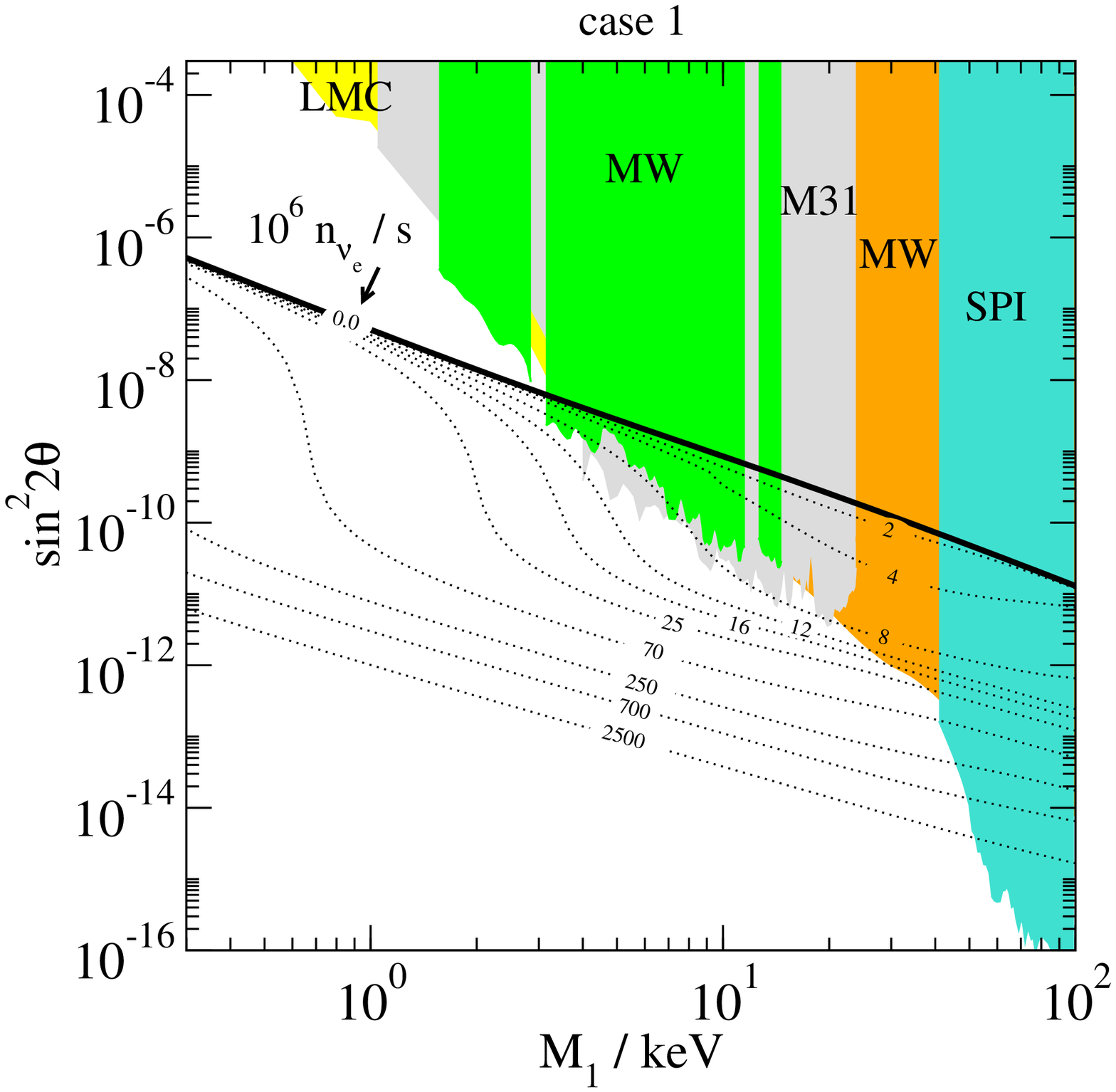}%
~~~\epsfysize=7.5cm\epsfbox{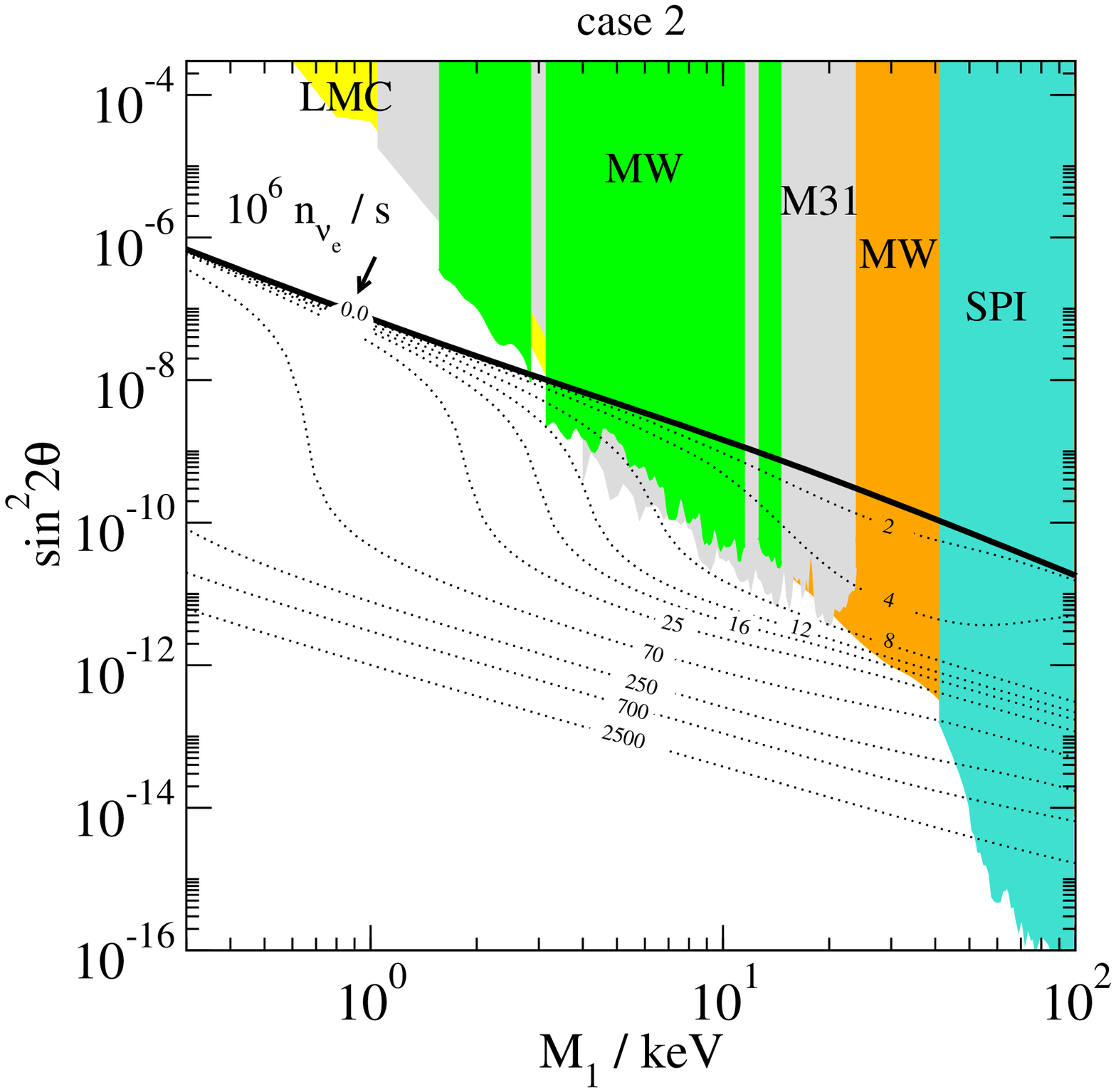}%
}

\caption[a]{\small 
The central region of \fig\ref{fig:exclusion_th},
$M_1 = 0.3 \ldots 100.0$~{keV}, compared with
regions excluded by various X-ray 
constraints~%
\cite{Boyarsky:2006fg,Boyarsky:2006ag,Boyarsky:2007ay,Boyarsky:2007ge},
coming from  XMM-Newton observations of 
the Large Magellanic Cloud (LMC), the Milky Way (MW), 
and the Andromeda galaxy (M31).  SPI marks the constraints from 
5 years of observations of the Milky Way galactic center by 
the SPI spectrometer on board the Integral observatory.} 
\la{fig:exclusion}
\end{figure}

The theoretical upper bound from \eq\nr{Ya1_bound} is compared with
experimental  constraints (from the non-observation of any X-ray
sterile neutrino decay peak in various presumed dark matter
concentrations) in  \fig\ref{fig:exclusion}.  A more detailed
discussion  concerning the implications of \fig\ref{fig:exclusion}
follows in \se\ref{se:astro}. 

Finally, we note that 
the plots in this section were produced without taking 
into account the back reaction discussed in \se\ref{ss:BR2}, 
i.e., by using the distributions $\f_\mp^{(0)}$. 
By recomputing $Y_{\alpha 1}$ for a number of masses 
and asymmetries from the corrected distributions $\f_\mp$, 
we find that the errors are maximal, $\sim 25\%$,  
for small masses, $M_1 \lsim 3$~keV, and intermediate 
asymmetries, $n_{\nu_e}/s \sim 10 - 20 \times 10^{-6}$. 
For larger masses and other asymmetries, the error from 
the omission of the back reaction is typically below 10\%, 
which we estimate to be well below our other systematic uncertainties.

%
\section{Sterile neutrino spectrum}
\la{se:spectrum}

We now move from the integrated total sterile neutrino  abundance,
\eq\nr{na1}, to the momentum distribution function, 
\eq\nr{distribution}. The physics context where this plays a role is
that of structure formation, particularly at the  smallest scales
(Lyman-$\alpha$ data). The corresponding constraints are considered
to be subject to more uncertainties than the X-ray bounds, both as
far as direct observational issues are concerned, as well as with
regard to dark matter simulations,  which have not been carried out
with actual non-equilibrium spectra so far. Nevertheless,
adopting a simple recipe for estimating the non-equilibrium effects
(cf.\ \eq\nr{M0_c}), the results of
refs.~\cite{Seljak:2006qw,Viel:2006kd} can be re-interpreted as the
constraints $\Ms \gsim 11.6$ keV and $\Ms \gsim 8$ keV,  respectively
(95\% CL), at vanishing asymmetry~\cite{als2}. Very recently limits
stronger by a factor 2--3 have been  reported~\cite{Viel:2007mv}.  
We return to how the constraints change in the case of a non-zero lepton 
asymmetry in \se\ref{se:astro}. We note, however, that the most 
conservative bound,  the so-called Tremaine-Gunn
bound~\cite{Tremaine:1979we,Lin:1983vq},   is much weaker and reads
$\Ms \gsim 0.3$ keV \cite{Dalcanton:2000hn},  which we have chosen as
the lower end of the horizontal axes in \figs\ref{fig:exclusion},
\ref{fig:avq}.

\begin{figure}[t]


\centerline{%
\epsfysize=7.5cm\epsfbox{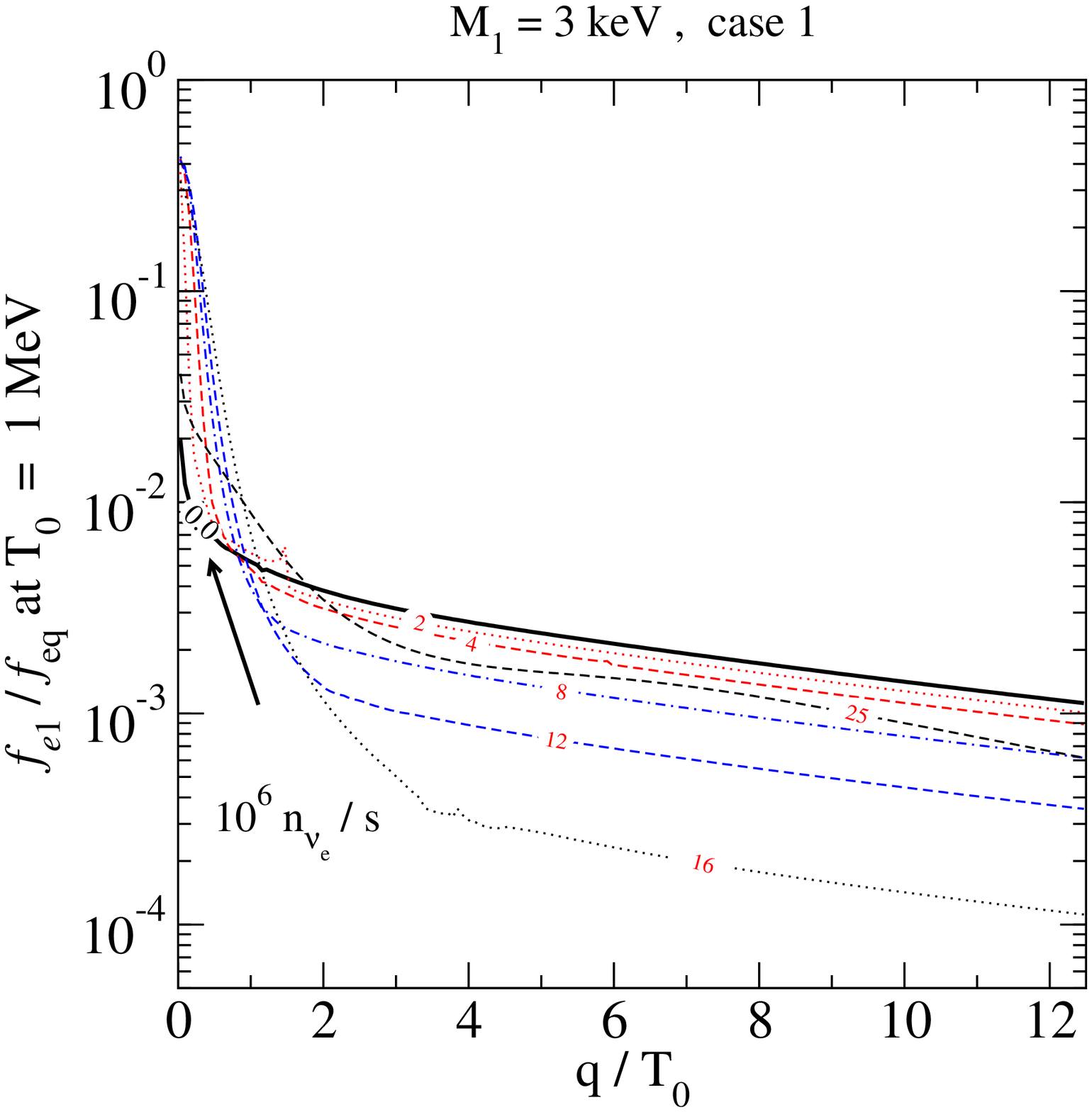}%
~~~\epsfysize=7.5cm\epsfbox{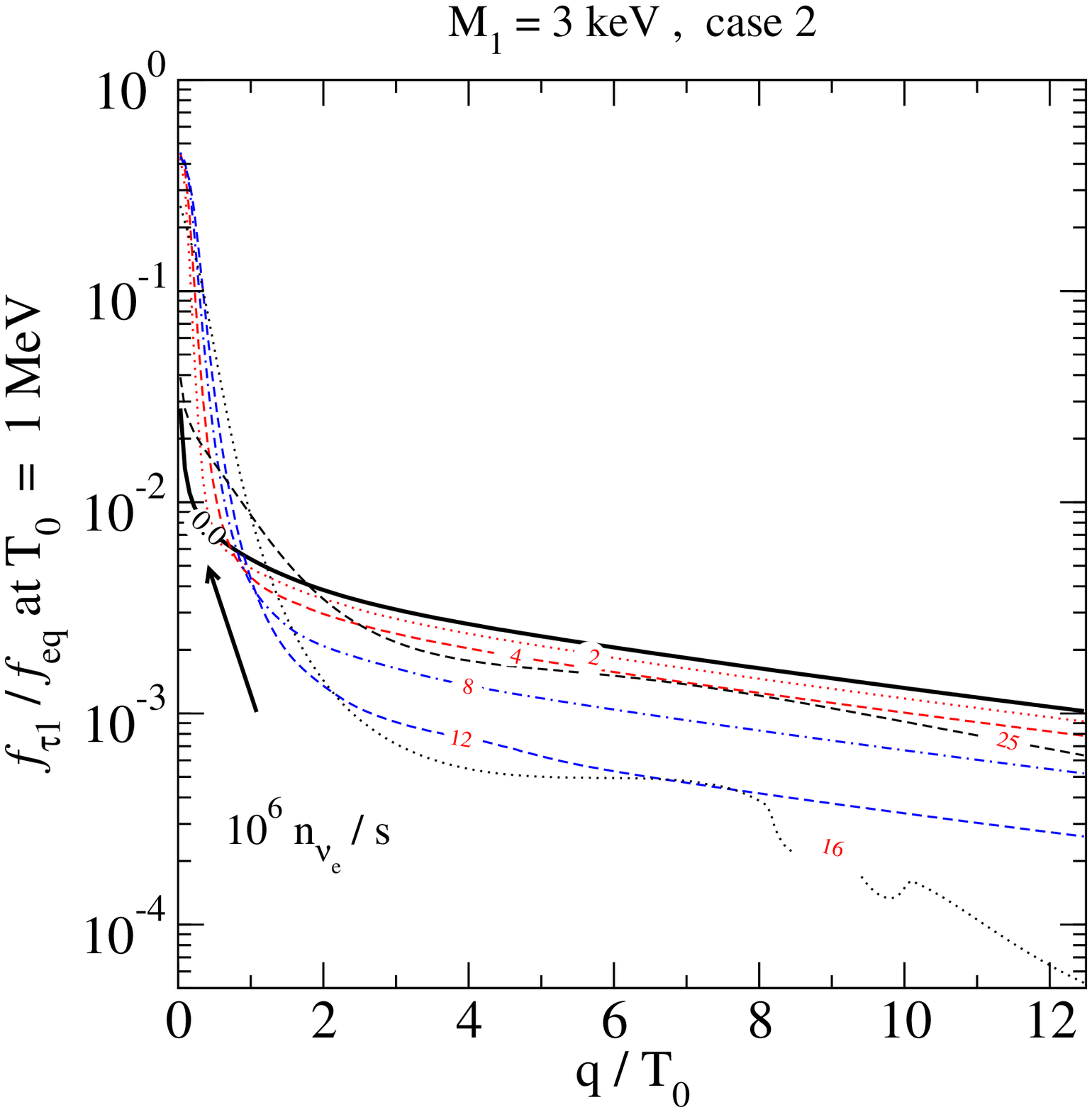}%
}

\caption[a]{\small  
The distribution function $\f_{\alpha 1}(t_0,q)$,
for $T_0 = 1$~MeV and  $M_1 = 3$~keV, normalised to the 
massless equilibrium value, $\f_\rmi{eq}(t_0,q) = 2 \nF{}(q) /(2\pi)^3$.
Left: case 1.
Right: case 2.
These results can be compared with 
refs.~\cite{Shi:1998km,Abazajian:2001nj}: the general 
feature of strong enhancement at small momenta is the same, 
but our distribution functions show more structure. The case
$n_{\nu_e}/s = 16\times 10^{-6}$ is particularly complicated
(and sensitive to uncertainties), since the resonance happens
to lie just on top of the QCD crossover, at $T\sim 150-200$~MeV, 
cf.\ \figs\ref{fig:Tevol}--\ref{fig:exclusion_th}.
} 
\la{fig:distr}
\end{figure}

\begin{figure}[t]


\centerline{%
\epsfysize=7.5cm\epsfbox{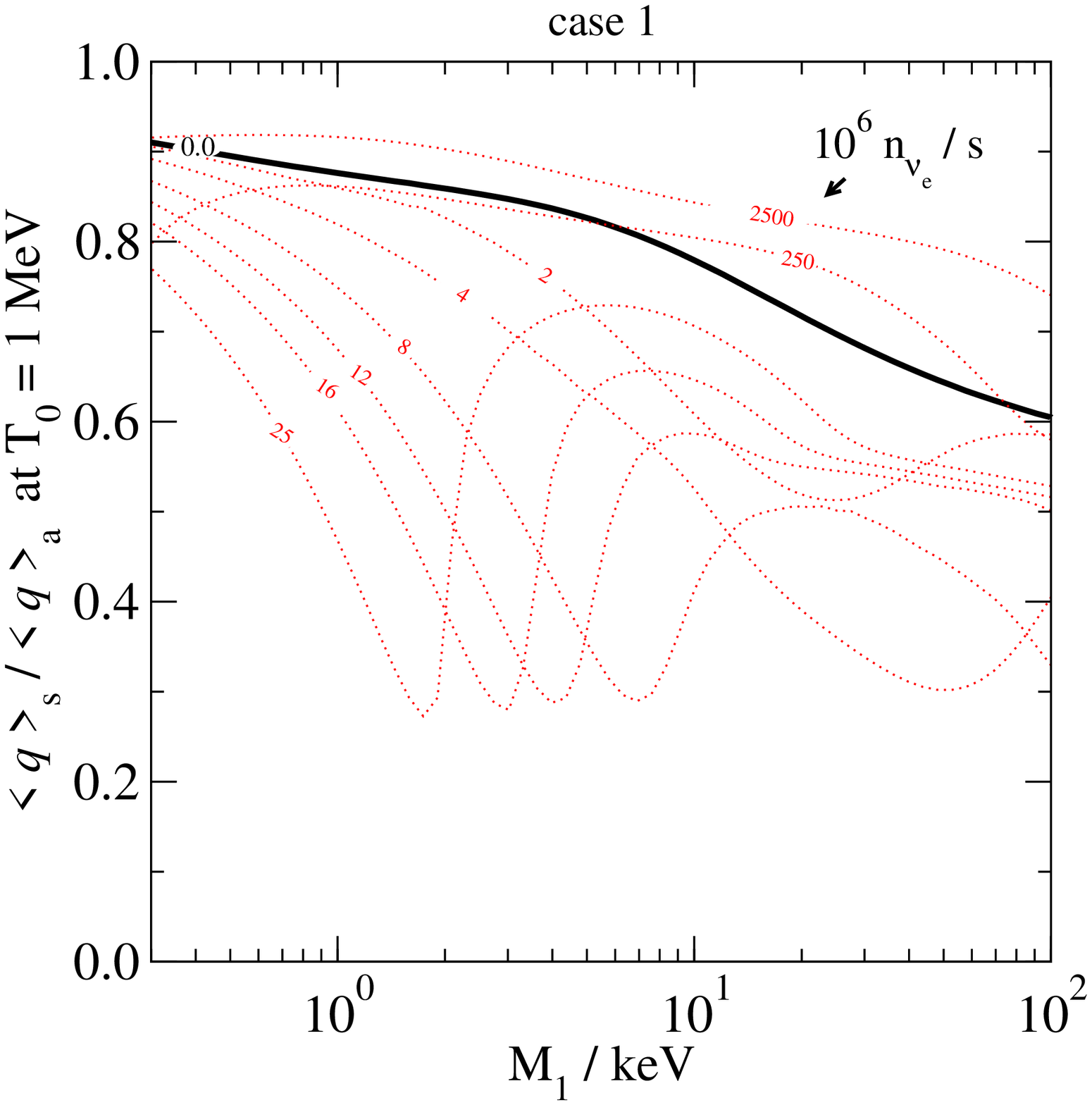}%
~~~\epsfysize=7.5cm\epsfbox{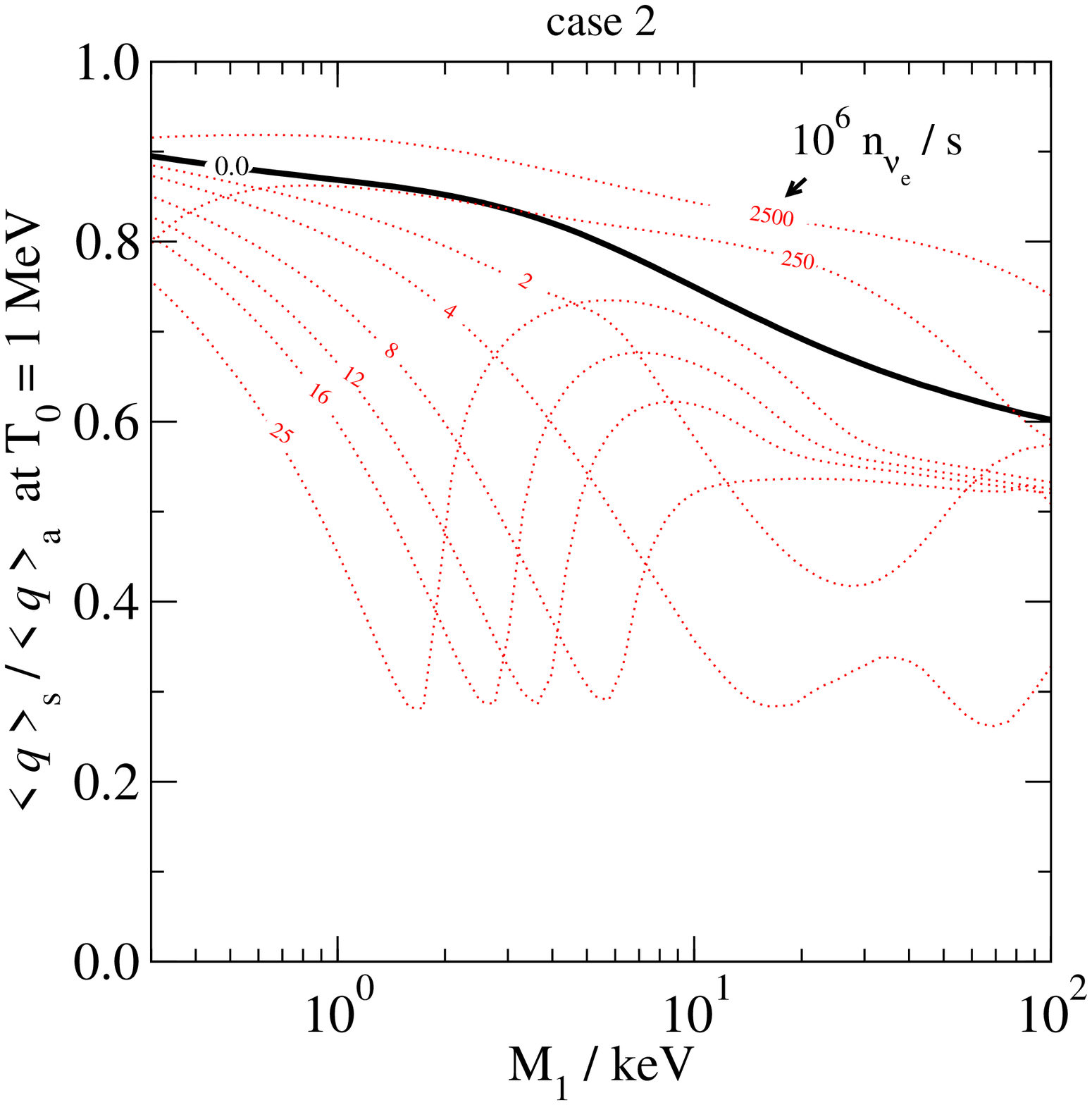}%
}

\caption[a]{\small  The average sterile neutrino momentum, 
$
 \langle q \rangle_{s} 
$,
normalised to the active neutrino 
equilibrium value,
$
 \langle q \rangle_{a} \equiv 7 \pi^4 T_0/180 \zeta(3) 
 \approx 3.15T_0
$. 
Left: case 1.
Right: case 2.
} 
\la{fig:avq}
\end{figure}

In \fig\ref{fig:distr} we show  examples of the spectra,  for a
relatively small mass $M_1 = 3$~keV (like in \fig\ref{fig:Tevol}), 
at which point the significant changes caused by the asymmetry can be
clearly identified.  The general pattern to be observed in
\fig\ref{fig:distr} is that for a small asymmetry, the distribution
function is boosted  only at very small momenta.  Quantities like the
average momentum $\langle q \rangle_s$  then decrease, as can be seen
in  \fig\ref{fig:avq}. For large asymmetry, the resonance affects all
$q$;  the total abundance is strongly enhanced with respect to the
case without a resonance, but the shape of the distribution function
is less distorted than at small asymmetry, so that the average momentum
$\langle q \rangle_s$ returns back towards the value in the 
non-resonant case. Therefore, for any given mass, we can observe  a
minimal value of $\langle q \rangle_s$ in \fig\ref{fig:avq}, 
$\langle q \rangle_s \gsim 0.3 \langle q \rangle_a$. This minimal
value is remarkably independent of $M_1$,  but the value of asymmetry
at which it is reached decreases with increasing $M_1$.  

Let us stress, however, 
that the values in \fig\ref{fig:avq} were produced without taking 
into account the back reaction discussed in \se\ref{ss:BR2}, 
i.e., by using the distributions $\f_\mp^{(0)}$. 
By recomputing $\langle q \rangle_{s} / \langle q \rangle_{a}$ 
for a number of masses and asymmetries from the approximatively 
corrected distributions $\f_\mp$ (shown in \fig\ref{fig:distr}), 
we find that for small masses, $M_1 \lsim 3$~keV, and 
intermediate asymmetries, $n_{\nu_e}/s \sim 10 - 20 \times 10^{-6}$, 
the results in \fig\ref{fig:avq} may be
{\em too small} by up to $\sim 25\%$. Thus, for small masses,  
the minimal average momentum may be better approximated
as $\langle q \rangle_s \gsim 0.4 \langle q \rangle_a$.
For larger masses and other asymmetries, the error from 
the omission of the back reaction is typically below 10\%, 
which we estimate to be below our other systematic uncertainties.

%
\section{Astrophysical constraints}
\la{se:astro}

The purpose of the present section is to combine the results of the
previous two sections, and derive the astrophysical constraints that
follow from them. Let us start by briefly recapitulating, once more, 
the two different types of considerations that we have carried out. 

First of all,  the theoretical computation described in
\se\ref{se:abundance} produces, for a given mass $M_1$ and mixing
angle $\sin^2\! 2 \theta$,  a definite total abundance of sterile
neutrinos. Requiring that this  abundance account for all of the
observed energy density in  dark matter, leads to the (lepton
asymmetry dependent)  mass--angle relation shown in
\fig\ref{fig:exclusion}.

The most direct constraint on the sterile neutrino dark matter 
scenario comes from comparing these curves with X-ray observations
(\fig\ref{fig:exclusion}).  For $n_{\nu_e}/s = 0.0$ we are in the
allowed region only for $M_1 \le 3$ keV. Increasing the asymmetry to
$n_{\nu_e}/s \simeq 8 \times 10^{-6}$ opens suddenly a whole
range of allowed mass values, up to $M_1 \simeq 25$ keV. Increasing 
the asymmetry further relaxes the upper bound even more but 
rather slowly; for instance,  if the
asymmetry is increased to
$n_{\nu_e}/s \simeq 25 \times 10^{-6}$, then we read from
\fig\ref{fig:exclusion} the upper bound $M_1 \lsim 40$ keV, while
the maximal allowed asymmetry $n_{\nu_e}/s \simeq 2500 \times 10^{-6}$ 
yields the  upper bound $M_1 \lsim 50$ keV.

Another important effect comes from the 
modification of the sterile neutrino spectrum through a lepton
asymmetry. As already found 
in ref.~\cite{Shi:1998km}, the non-equilibrium spectrum of the dark 
matter sterile neutrinos created in the presence of a lepton asymmetry 
is very different from the thermal one. Some examples are shown 
in \fig\ref{fig:distr}.

Now, an observation of small scale structures in the  Lyman-$\alpha$
data puts an upper bound on the free-streaming length  and,
consequently, on the average velocity of the dark matter particles. 
This converts to a lower bound on the inverse velocity, which, in
the  absence of an actual analysis with non-equilibrium spectra,  can
be roughly estimated as~\cite{Hansen:2001zv} 
\be 
 M_1 \frac{\langle {q} \rangle_a }{ \langle {q} \rangle_s } \gsim M_0
 \quad
  \Leftrightarrow
 \quad
 M_1 \gsim M_0 \frac{\langle {q} \rangle_s }{ \langle {q} \rangle_a }
 \;, \la{M0_c}
\ee
where $\langle {q} \rangle_a$  and $\langle {q} \rangle_s$ are the
average momenta of active and sterile neutrinos, respectively,   at
the moment of structure formation, and the value of $M_0$ is  $M_0
\simeq 14$ keV (95\% CL) according to ref.~\cite{Seljak:2006qw} (or
$M_0 \simeq 10$ keV at 99.9\% CL). According to
ref.~\cite{Viel:2006kd} the bound is  somewhat weaker, while
according to ref.~\cite{Viel:2007mv} it could be as strong as $M_0
\simeq 28$ keV (95\% CL). Let us start by considering the most conservative
bound  $M_0=10$ keV.\footnote{%
 In this work we only consider lower bounds on the mass 
 of the dark matter particle from structure formation. 
 However, the problems of missing satellites and cuspy
 profiles in Cold Dark Matter cosmological models, 
 as well as that of the galactic angular momentum, 
 suggest that an upper bound may exist as well~\cite{cusp}.
}

The dependence of $\langle {q} \rangle_s / \langle {q} \rangle_a$
on $M_1$ and the lepton asymmetry is shown in \fig\ref{fig:avq}. 
Quite interestingly, this ratio can decrease to $0.3$
(or $0.4$ at small $M_1$, cf.\ end of \se\ref{se:spectrum})
for a certain range of asymmetries. However, 
$\langle {q} \rangle_s / \langle {q} \rangle_a$ does not decrease 
further with increasing asymmetry, but increases again. 
Therefore, the lower limit $M_0 = 10$~keV corresponds
to $M_1 \gsim 4$~keV.

Combining now the two constraints (from X-rays,  
\fig\ref{fig:exclusion}, and from structure formation, \eq\nr{M0_c}), 
we observe that a solution satisfying both constraints exists 
for $n_{\nu_e}/s \gsim 8 \times 10^{-6}$. The 
solution corresponds to masses $M_1 \simeq 4 - 25$ keV and 
mixing angles $\sin^2\! 2\theta \simeq 8 \times 10^{-10} - 
2 \times 10^{-12}$.  If the lepton asymmetry is increased, larger 
masses and smaller mixing angles become possible.   

Let us then consider the case $M_0 \simeq 28$ keV~\cite{Viel:2007mv}.
Using the minimal value 
$\langle q\rangle_s/\langle q\rangle_a \sim 0.3$, 
$M_0 \simeq 28$ keV corresponds to
$M_1 \gsim 28 \times 0.3 \simeq 8.4$ keV
according to \eq\nr{M0_c}. 
Combining this with the X-ray constraints in \fig\ref{fig:exclusion},
we see that this can in fact be satisfied with the same asymmetry
as before, $n_{\nu_e}/s \gsim 8 \times 10^{-6}$. 

It should be stressed, however, that the validity of \eq\nr{M0_c} for
spectra as extreme as those in \fig\ref{fig:distr} remains to  be
cross-checked. Nevertheless, if true, we can establish the absolute 
lower bound $M_1 \ge 4$~keV for sterile neutrinos capable of 
accounting for all of the dark matter in the Universe (assuming that
the only mechanism for dark matter sterile neutrino production is
active-sterile mixing).

%
\section{Conclusions}
\la{se:concl}

The $\nu$MSM, i.e.\ Minimal Standard Model extended by three
right-handed neutrinos with masses smaller than the electroweak 
scale, has a number of parameters not appearing in  the Standard
Model: three Majorana masses and a $3\times 3$ complex matrix of
Yukawa couplings.  In ref.~\cite{asy}, the part of the parameter
space  associated with the two heaviest right-handed neutrinos  was
explored in detail, and a phenomenologically interesting corner was
identified. Specifically,  it was found that if the mass difference
of the heavy (dominantly right-handed) neutrino mass eigenstates is 
much {\em smaller} than the known mass differences of the light 
(dominantly left-handed) mass eigenstates ({\bf Scenario IIa}), 
then it is possible to explain the known active neutrino
mass differences and mixings, and simultaneously generate and
subsequently maintain a significant lepton asymmetry in the
Universe,  without violating constraints related to Big Bang
Nucleosynthesis  at temperatures of about 0.1 MeV. 

The purpose of the present paper has been to constrain the parameters
associated with the lightest of the right-handed neutrinos,  referred
to with the subscript ``1''. In this case there  are no constraints
from the known  active neutrino mass  differences and mixings;
rather, the contribution from  the lightest right-handed neutrinos to
the see-saw formulae  is much below 0.01~eV. (Consequently the $\nu$MSM
excludes the case of degenerate active neutrinos with a common mass
scale  $\gg 0.01$~eV; in particular, the effective mass for neutrinoless
double beta decay cannot exceed 0.05~eV~\cite{Bezrukov:2005mx}.) 
In contrast, the assertion that all of 
dark matter be made of these neutrinos does allow us to place
further constraints on the parameters. More precisely, we were led
in \se\ref{se:astro} to the mass range $M_1 \simeq 4 \ldots 50$~keV.  The
lower bound originates from combining the theoretical analysis of the
present paper with  observational X-ray (\fig\ref{fig:exclusion}) 
and structure formation (\eq\nr{M0_c}, \fig\ref{fig:avq}) constraints, 
whereas the
upper bound is dictated by the maximal lepton asymmetry allowed
by Big Bang Nucleosynthesis.  
The absolute values of the  Yukawa
couplings of the lightest right-handed neutrinos should be in the
range  $5\times 10^{-15} \ldots 4 \times 10^{-13}$ in this case.  

Of course, these
constraints are relaxed if the sterile neutrinos only account for a
fraction of the dark matter (see, e.g., ref.~\cite{Palazzo:2007gz}); 
if a part of them are produced by
some non-equilibrium mechanism not related to active-sterile 
mixing (see, e.g., 
refs.~\cite{Shaposhnikov:2006xi,Kusenko:2006rh,Petraki:2007gq}); 
or if the thermal history of the Universe is non-standard
(see, e.g., refs.~\cite{Gelmini:2004ah,Yaguna:2007wi,Khalil:2008kp}).

It is important to stress, in addition, that the lower bound 
$M_1 \gsim 4$~keV relies on a naive re-interpretation of structure
formation simulations which were carried out assuming  a thermal
spectrum of dark matter particles, rather than  a proper
non-equilibrium shape as given in \fig\ref{fig:distr}.  Hopefully
this issue can be put on more solid ground soon. 

Finally, we recall that perhaps the most realistic hope for an  
experimental detection of dark matter sterile neutrinos in the
parameter range that we have discussed,  would be through the
discovery of a peak in the diffuse X-ray background from regions
where dark matter decays. The dominant decay  channel is $N_1 \to
\nu\gamma$ and the spectrum should thus  peak at the energy 
$M_1/2 \gsim 2$~keV.  As far as laboratory searches are concerned, they are
quite  difficult due to the  very small Yukawa couplings of the  dark
matter sterile neutrinos; however, a possibility  has been suggested
in ref.~\cite{Bezrukov:2006cy}. 

%
\section*{Acknowledgements}

We thank Alexey Boyarsky and 
Oleg Ruchayskiy for providing  the X-ray data plotted in
\fig\ref{fig:exclusion} and for helpful remarks. 
The work of M.L. was supported in part  by the National Science Foundation, 
under Grant No.\ PHY05-51164,   
and that of M.S.\ by the Swiss National Science Foundation.
We thank Takehiko Asaka for collaboration at initial stages of this
work.



\appendix
\renewcommand{\thesection}{Appendix~\Alph{section}}
\renewcommand{\thesubsection}{\Alph{section}.\arabic{subsection}}
\renewcommand{\theequation}{\Alph{section}.\arabic{equation}}

%
\section{Different characterizations of lepton asymmetry}
\la{se:asy}

Several different conventions are used in the literature for 
characterizing the presence of a non-zero lepton asymmetry, and for 
completeness we specify the relations between them here. 
In analogy with the characterization of the baryon asymmetry through $n_B/s$,
the conceptually cleanest way is to give the ratio of 
lepton asymmetry  density over the total entropy density, 
because this quantity remains constant as a function of the temperature 
(as long as the Universe remains in thermodynamic equilibrium). 
However there are many leptonic species, so we need to specify which
ones to count; in this paper we have assumed that the asymmetries are
equal in all active leptons (both left-handed and right-handed), 
and choose the electron-like neutrinos (two degrees of freedom, 
particles and antiparticles) as a representative. The corresponding
asymmetry, ``particles minus antiparticles'',
normalised to the entropy density, 
is then denoted by $n_{\nu_e}/s$. 

Another way to express the asymmetry is to give the leptonic chemical 
potential, $\mu_L$, that corresponds to $n_{\nu_e}$. This is useful, 
since chemical potential is the quantity that appears in quantum
field theoretic computations. In the free limit, 
\be
 n_{\nu_e} = \langle \hb{\nu}_{eL} \gamma_0 \hat \nu_{eL} \rangle 
 = 2 i \left. \Tint{Q_\fe} \frac{\tilde q_0}{\tilde Q^2 + m^2}
 \right|_{\tilde q_0 = \omega_\fe + i \mu_L}
 = \int\! \frac{{\rm d}^3\vec{q}}{(2\pi)^3}
 \Bigl[ \nF{}(E-\mu_L) - \nF{}(E+\mu_L)\Bigr]
 \;, \la{neLfree}
\ee
where $Q_\fe$ are the fermionic Matsubara momenta, 
$Q_\fe \equiv (\omega_\fe,\vec{q})$ with 
$\omega_\fe \equiv 2 \pi T(n + \fr12)$, $n \in \ZZ$; 
$\tilde Q \equiv (\tilde q_0,\vec{q}) \equiv Q_\fe + (i\mu_L,\vec{0})$;
and $E \equiv \sqrt{\vec{q}^2 + m^2}$. 
In the massless limit $m \ll T$, \eq\nr{neLfree} evaluates to 
\be
 n_{\nu_e} = \frac{\mu_L T^2}{6} + \frac{\mu_L^3}{6\pi^2}
 \;.
\ee
Therefore, for $\mu_L \ll T$, 
\be
 \frac{n_{\nu_e}}{s} \approx \frac{15}{4\pi^2 h_\rmi{eff}}
 \times \frac{\mu_L}{T}
 \;, \la{n_vs_mu}
\ee
where $h_\rmi{eff}$ parametrizes the total entropy density through
$
 s \equiv {2 \pi^2 T^3} h_\rmi{eff} / {45} 
$.

Yet another way to characterize the asymmetry
is to use the ``intuitive'' measure 
of the asymmetry that one is naturally lead to  
in Boltzmann-equation or density-matrix type treatments, 
where the basic degrees of freedom are on-shell particle states:
\be
 \Delta  
 \equiv \frac{\#_{\nu_e} - \#_{\bar\nu_e}}{\#_{\nu_e} + \#_{\bar\nu_e}}
 = \frac{n_{\nu_e}}{n_\rmi{eq}}
 \;, \la{Delta_def} 
\ee 
where 
$
 n_\rmi{eq} \equiv \#_{\nu_e} + \#_{\bar\nu_e}
 \equiv 2 \int {\rm d}^3\vec{q} /(2\pi)^3/(e^{q/T}+1) = 3\zeta(3)T^3/2\pi^2
$.
Then we get 
\be
 \frac{n_{\nu_e}}{s} = \frac{135\zeta(3)}{4\pi^4 h_\rmi{eff}}
 \times \Delta
 \;.
\ee
Or, one can normalize with respect to the photon density, 
\be
 L \equiv \frac{n_{\nu_e}}{n_\gamma}
 \;, 
\ee
where 
$
 n_\gamma
 \equiv 2 \int {\rm d}^3\vec{q} /(2\pi)^3/(e^{q/T}-1) = 2 \zeta(3)T^3/\pi^2
$, 
which yields 
\be
 \frac{n_{\nu_e}}{s} = \frac{45\zeta(3)}{\pi^4 h_\rmi{eff}}
 \times L 
 \;.
\ee
Numerically, choosing to cite the temperature-dependent 
asymmetries ($\mu_L/T, \Delta, L$) at $T = 100$~GeV as e.g.\ in 
refs.~\cite{Abazajian:2002yz,Abazajian:2006yn}, 
we may approximate $h_\rmi{eff}\approx 102$ 
for the MSM,\footnote{
  Refs.~\cite{pheneos,als2} give estimates for $h_\rmi{eff}$ from this range
  down to low temperatures, $T \sim 1$~MeV, taking into account radiative
  corrections due to strong interactions.} 
whereby 
\ba
 \frac{n_{\nu_e}}{s}
 & \approx &  
 3.7 \times 10^{-3} \times \left. \frac{\mu_L}{T} 
 \right|_\rmi{$T = 100$~GeV}
 \\ & \approx & 
 4.1 \times 10^{-3} \times \left. \Delta
 \right|_\rmi{$T = 100$~GeV}
 \\ & \approx &  
 5.4 \times 10^{-3} \times \left. L 
 \right|_\rmi{$T = 100$~GeV}
 \;.
\ea



\end{document}